\documentclass[letterpaper,11pt]{article}
\pdfoutput=1
\usepackage{jheppub}
\usepackage{epstopdf}
\usepackage{slashed}
\usepackage{graphicx} 
\usepackage{subfig} 
\usepackage{soul}
\usepackage{amsmath}
\usepackage{multirow}
\usepackage{tablefootnote}
\usepackage{comment}
\usepackage{color}
\usepackage[utf8]{inputenc}
\usepackage{subfig}

\usepackage{graphicx}

\usepackage{amsmath}
\usepackage{amsfonts}
\usepackage{amssymb}
\usepackage{cleveref}
\usepackage{color}
\usepackage{enumitem}

\def\be{\begin{equation}}
\def\ee{\end{equation}}
\def\bea{\begin{eqnarray}}
\def\eea{\end{eqnarray}}

\def\vep{\varepsilon}

\def\<{\langle}
\def\>{\rangle}

\def\mt{{\mathcal{T}}}

\def\vep{\varepsilon}

\def\mt{{\mathcal{T}}}

\def\hc{\text{h.c.}}

\def\slashchar#1{\setbox0=\hbox{$#1$}           
   \dimen0=\wd0                                 
   \setbox1=\hbox{/} \dimen1=\wd1               
   \ifdim\dimen0>\dimen1                        
      \rlap{\hbox to \dimen0{\hfil/\hfil}}      
      #1                                        
   \else                                        
      \rlap{\hbox to \dimen1{\hfil$#1$\hfil}}   
      /                                         
   \fi}          

\def\mD{\mathcal{D}}

\def\mI{\mathcal{I}}

\def\mL{\mathcal{L}}

\def\mO{\mathcal{O}}

\def\mT{\mathcal{T}}

\def\bea{\begin{eqnarray}}
\def\eea{\end{eqnarray}}
\def\beq{\begin{equation}}
\def\eeq{\end{equation}}
\def\vep{\varepsilon}

\def\<{\langle}
\def\>{\rangle}

\def\ewd{{\mathcal{D}}}

\newcommand{\Eq}[1]{Eq.~(\ref{#1})}
\newcommand{\Ref}[1]{Ref.~\cite{#1}}

\newcommand{\olr}[1]{\overleftrightarrow{#1}}

\preprint{CERN-TH-2018-174}

\title{Universal Relations in Composite Higgs Models}

\author[a]{Da Liu,}
\author[a,b]{Ian Low}
\author[b]{and Zhewei Yin}

\affiliation[a]{High Energy Physics Division, Argonne National Laboratory, Argonne, IL 60439, USA}
\affiliation[b]{Department of Physics and Astronomy, Northwestern University, Evanston, IL 60208, USA}

\emailAdd{da.liu@anl.gov}
\emailAdd{ilow@anl.gov}
\emailAdd{zheweiyin2015@u.northwestern.edu}

\abstract{
We initiate a phenomenological study of ``universal relations'' in composite Higgs models, which  are  dictated by nonlinear shift symmetries acting on the 125 GeV Higgs boson. These are relations among one Higgs couplings with two electroweak gauge bosons (HVV), two Higgses couplings with two electroweak gauge bosons (HHVV), one Higgs couplings with three electroweak gauge bosons (HVVV), as well as triple gauge boson couplings (TGC), which are all controlled by a single input parameter: the decay constant $f$ of the pseudo-Nambu-Goldstone Higgs boson.  Assuming custodial invariance in strong sector, the  relation is independent of the symmetry breaking pattern in the UV, for an arbitrary  symmetric coset $G/H$. The complete list of corrections to HVV, HHVV, HVVV and TGC couplings in composite Higgs models is presented to all orders in $1/f$, and up to four-derivative level,  without referring to a particular $G/H$. We then present several examples of  universal relations in ratios of coefficients which could be extracted experimentally. Measuring the universal relation requires a precision sensitive to effects of dimension-8 operators in the effective Lagrangian and highlights the importance of verifying the tensor structure of HHVV interactions in the standard model, which remains untested to date.

}

\begin{document}

\maketitle

\noindent
\section{Introduction}
\label{sect:Intro}

Measurements of properties of the 125 GeV Higgs boson at the Large Hadron Collider (LHC) so far  focus on processes involving one Higgs boson with two other SM particles, such as gauge bosons or heavy-flavor quarks \cite{Khachatryan:2016vau}. The consistency of these measurements  with predictions from  the standard model (SM) gives confidence to the ``Higgs nature'' of the 125 GeV boson. A much more open question is whether this is {\em the} SM Higgs boson, whose interactions are completely determined by its mass and other SM input parameters. The best precision in current measurement lies in the HVV couplings, which is of the order of 10\%. Recall that effects of new particles with a mass of 1 TeV or higher are generically of the order of 5\% or less. Therefore, it is perhaps not surprising that no credible deviation has shown up to date. Nevertheless, the study on HVV couplings is very thorough, and involves not only the signal strength, but also the predicted tensor structure of the coupling \cite{Gainer:2011xz,Bolognesi:2012mm,Stolarski:2012ps,Artoisenet:2013puc,Gainer:2014hha}. On the other hand,  due to the limitation in the center-of-mass energy of the LHC, processes involving two Higgs bosons such as the Higgs trilinear coupling and HHVV couplings have very small production rates and remain as untested predictions of the SM. 

In spite of the tremendous amount of experimental efforts,  there  are still outstanding theoretical questions to be answered. In the SM the electroweak symmetry  breaking (EWSB) is triggered by the vacuum expectation value (VEV) of the Higgs doublet $H$, whose potential is 
\be
\label{eq:Higgspot}
V(H) = - \mu^2 H^\dagger H + {\lambda} |HH^\dagger|^2 \ .
\ee
The Mexican hat potential is reminiscent of the effective potential for conventional superconductivity  proposed by Ginzburg and Landau in 1950 \cite{Ginzburg:1950sr}. In both the SM and the Ginzburg-Landau theory, the crucial ``$-$" sign in front of the quadratic term in the potential is completely ad hoc, without a microscopic understanding.  In 1957, Bardeen, Cooper and Schrieffer (BCS) gave a {\em microscopic} theory of conventional superconductivity \cite{Bardeen:1957mv}, which allows one to compute the coefficients $\mu^2$ and $\lambda$ in the Ginzburg-Landau theory. To the contrary, we do not yet have a  microscopic theory for the Higgs potential, as well as the crucial minus sign, even to date. It is a somewhat shocking realization that, more than forty years after the Higgs boson was proposed in Ref.~\cite{Higgs:1964pj}, our understanding of the EWSB is still as primitive as the Ginzburg-Landau theory.

Of course, the lack of a BCS level understanding of EWSB  is not without trying. The direct analogy of the BCS theory, where the spontaneous symmetry breaking occurs dynamically, goes by the name of ``technicolor'' and is now strongly disfavored by experimental data \cite{Patrignani:2016xqp}. The compatibility with data can be improved by proposing additional spontaneously broken global symmetries above the electroweak scale, under which the 125 GeV Higgs arises as a pseudo-Nambu-Goldstone boson  \cite{ArkaniHamed:2001nc,ArkaniHamed:2002qx,ArkaniHamed:2002qy,Contino:2003ve,Agashe:2004rs}. In this scenario, the Mexican hat potential is generated radiatively \`{a} la the celebrated Coleman-Weinberg mechanism \cite{Coleman:1973jx}. This class of models is now referred to as the composite Higgs model and a survey of the literature reveals a garden variety of possibilities \cite{Bellazzini:2014yua}, each invoking a different coset structure $G/H$. Conventional wisdom, based on the seminal work of Callan, Coleman, Wess and Zumino (CCWZ) \cite{Coleman:1969sm, Callan:1969sn}, has it that different $G/H$ leads to a different low-energy effective Lagrangian. As a consequence, comparisons with the data are usually made on a model-by-model basis, with the minimal coset $SO(5)/SO(5)$ receiving the most attention \cite{Panico:2015jxa}.

In recent years it was realized that there is an infrared construction of the effective action of Nambu-Goldstone bosons that does not require prior knowledge of the symmetry breaking coset $G/H$ \cite{Low:2014nga,Low:2014oga}, by focusing on shift symmetries under which the Nambu-Goldstone boson transforms non-homogeneously and nonlinearly,
\be
\label{eq:shift}
\pi^a(x)\to \pi^a(x)+\epsilon^a +\cdots\ ,
\ee
where $\epsilon^a$ are constant and terms neglected are higher order and nonlinear in $\pi^a$. The IR construction turned out to be the algebraic realization, at the Lagrangian level \cite{Low:2017mlh,Low:2018acv},  of the ``soft bootstrapping" approach pursued by the amplitudes community to reconstruct effective theories using on-shell quantities \cite{Cheung:2014dqa,Cheung:2016drk,Rodina:2018pcb}. In the end, interactions of Nambu-Goldstone bosons only serve one purpose: producing the correct soft limit, the so-called Adler's zero \cite{Adler:1964um}, amid the constraint of unbroken, linearly realized symmetry. For example, for three Nambu-Goldstone bosons transforming as the adjoint of an unbroken $SU(2)$ subgroup of a possibly larger linearly realized symmetry group $H$, their self-interactions are entirely determined by producing the correct soft limit, subject to the constraints of the linearly realized $SU(2)$ subgroup, and remain agnostic to the rest of the coset structure $G/H$ except for the normalization of the decay constant $f$. For a viable composite Higgs models, the 125 GeV Higgs always transforms as the fundamental representation ($\mathbf{4}$) of an unbroken $SO(4)$ subgroup of $H$. As a result, their self-interactions, as well as interactions with electroweak gauge bosons, are universal in composite Higgs models, even after integrating out heavy composite resonances that are typically present in these models \cite{Liu:2018vel}.

Before we proceed further,  it is instructive to address three potential questions that may arise, especially from non-experts, regarding the universal relations:

\begin{enumerate}

\item {\bf Must the 125 GeV Higgs transform as the $\mathbf{4}$ of an $SO(4)$ subgroup of $H$?}\\
By construction the electroweak $SU(2)_L\times U(1)_Y$ group is broken only by the Higgs VEV and is unbroken above the weak scale. Thus the linearly realized symmetry $H$ must contain  $SU(2)_L\times U(1)_Y$ as a subgroup. However, it is well-known that the Higgs sector of the SM model contains an accidental $SO(4)\approx SU(2)_L\times SU(2)_R$ symmetry, referred to as the custodial invariance \cite{Sikivie:1980hm}, which is responsible for protecting the  precisely measured $\rho$ parameter. As a result, viable composite Higgs models typically choose an $H$ that contains the larger $SO(4)$ custodial symmetry. The compatibility with the $\rho$ parameter is much improved when the 125 GeV Higgs transforms as the $\mathbf{4}$ of $SO(4)$ subgroup.

\item {\bf Is the universality implied by the minimal coset $SO(5)/SO(4)$?}\\
Some might argue that, under the assumption of a single light Higgs boson at 125 GeV, the low-energy effective theory of an arbitrary coset $G/H$ must reduce to the minimal  $SO(5)/SO(4)$ coset, upon integrating out heavy composite resonances. The universality is therefore a consequence of the uniqueness of the minimal coset structure. It turns out that IR construction of the Nambu-Goldstone interactions in Refs.~\cite{Low:2017mlh,Low:2018acv} does not depend on the existence of other light degrees of freedom. For example, in models containing two light Higgs bosons,$H_1$ and $H_2$, the low-energy coset structure will not be $SO(5)/SO(4)$.\footnote{One example of composite two-Higgs-doublet models is the $SO(6)/SO(4)\times SO(2)$ coset studied in Ref.~\cite{Mrazek:2011iu}.} In this case the effective Lagrangian for self-interactions among $H_i$, $i=1,2$, will remain identical to the one from the $SO(5)/SO(4)$ coset, up to the normalization of the decay constant $f$.\footnote{Of course in this case there will be additional interactions between $H_1$ and $H_2$. In principle they can be determined by considering a larger set of shift symmetries.}

\item {\bf What is the impact of additional light degrees of freedom?} \\
Although the effective Lagrangian involving the self-interactions of the 125 GeV Higgs in composite Higgs models is universal, other light degrees of freedom, if there, could contribute to on-shell amplitudes of the Higgs boson as an intermediate propagator. Again using the two-Higgs-doublet-model as an example, a trilinear coupling like $h_1h_1h_2$ could arise after electroweak symmetry breaking, even for a symmetric coset.\footnote{For a symmetric coset $G/H$, one could impose an internal $Z_2$ symmetry which forbids trilinear couplings of Nambu-Goldstone bosons, although sometimes a slightly different variant of the $Z_2$ symmetry may also be implemented \cite{Cheng:2003ju,Cheng:2004yc,Low:2004xc}.} Therefore $h_2$ could contribute to S-matrix elements involving four external $h_1$ bosons. The existence of such a contribution does not invalidate the universal relations, which relate couplings in the effective Lagrangian, not S-matrix elements. However, their presence complicates the experimental effort to verify and test the universal relations, as additional observables and techniques  might be necessary to disentangle the H$_1$VV and H$_1$H$_1$VV couplings from the rest.  This is not dissimilar to efforts to measure different tensor structures of HVV couplings using differential distributions of decay products and multivariate techniques.

\end{enumerate}

This work is organized as follows. In Section \ref{sect:2} we begin with a non-technical argument leading to the infrared construction of Nambu-Goldstone effective actions without reference to a coset structure $G/H$. Then we provide a summary of the results in   general  and then specialize to the case of the Higgs transforming as the $\mathbf{4}$ of $SO(4)$. In Section \ref{sect:3} we give the complete list of operators contributing to HVV, HHVV, HVVV and TGC predicted in the composite Higgs models, up to all orders in $1/f$ and four-derivative level. Seven universal relations are presented, in the unitary gauge, relating the various coefficients which could be extracted experimentally. A preliminary phenomenological study on the universal relations is given in Section \ref{sect:4}, followed by the Conclusion. We also include two appendices:  Appendix \ref{appen:so5} on our convention of $SO(4)$ group generators and Appendix \ref{app:d6} on matching the universal nonlinear Lagrangian to linearized dimension-6 operators.

\noindent
\section{The Infrared Perspective}
\label{sect:2}

\subsection{An Overview}

Properties of Nambu-Goldstone bosons were studied intensively in the context of pions in low-energy QCD. A large body of work on ``soft pions'' exists in the literature, some of which turn out to be independent of the pattern of chiral symmetry breaking in QCD. One example that is especially relevant to our discussion is the Adler's zero condition \cite{Adler:1964um}, which states that on-shell amplitudes of pion scattering in the exact massless limit must vanish as one external momentum taken to zero, as a consequence of the spontaneously broken chiral symmetry.

\begin{figure}[t]
 \centering{ \includegraphics[height=4.5cm]{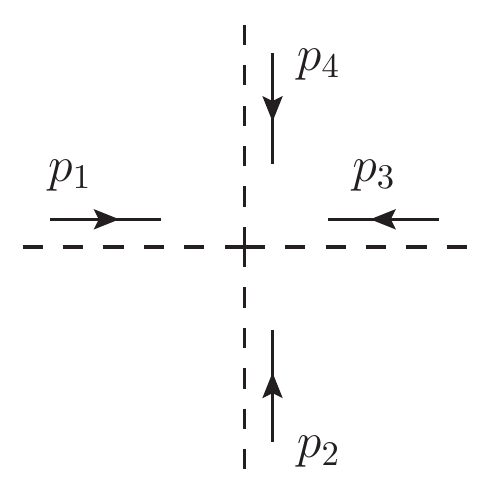}}
\caption{The 4-pt amplitude involves a single 4-pt interaction vertex, since there is no three-pt vertex among Nambu-Goldstone bosons for a symmetric coset. Adler's zero condition requires $M_4=s_{24}/f^2$.
}\label{fig:4pt}
\end{figure}

\begin{figure}[t]
\centerline{\includegraphics[height=3.2cm]{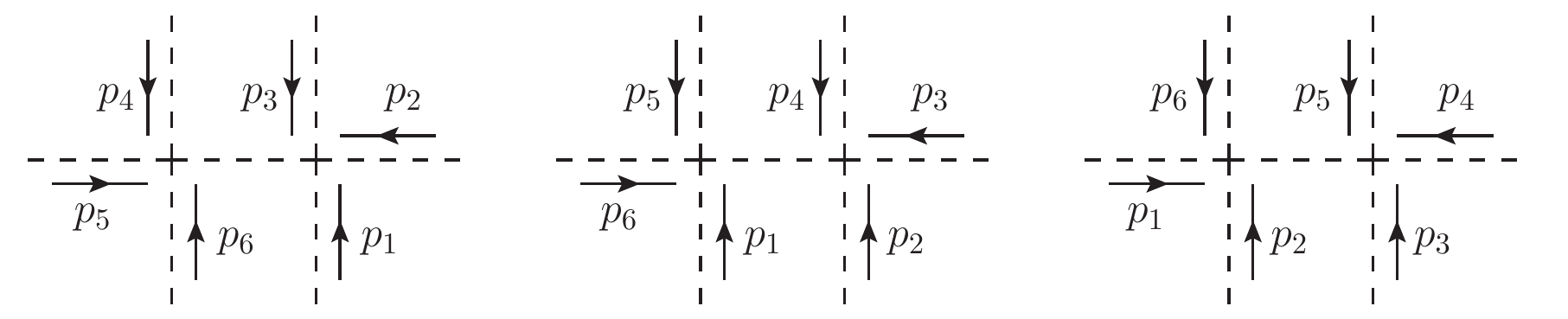}}
\caption{Contributions to $M_6$ from the 4-pt vertices: $(s_{13}s_{46}/P^2_{123}+s_{24}s_{15}/P^2_{234}+s_{35}s_{26}/P^2_{345})/f^2$.
}\label{fig:6pt_part}
\end{figure}

\begin{figure}[t]
\centerline{\includegraphics[height=3.5cm]{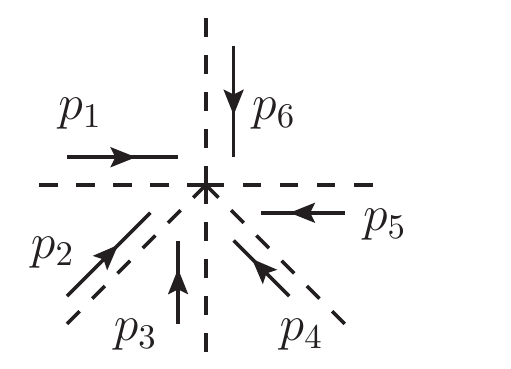}}
\caption{The 6-pt interaction vertex required by imposing the Adler's zero condition on $M_6$: $-P^2_{135}/f^2$.
}\label{fig:6pt_ver}
\end{figure}

It turns out that the Adler's zero condition can be taken as the defining property of Nambu-Goldstone bosons. By constructing S-matrix elements that always satisfy the Adler's zero condition, it is possible to arrive at the complete (tree-level) amplitudes of pions without referring to the notion of ``spontaneous symmetry breaking.'' At a practical level, the construction starts at the 4-point(pt) {\em partial amplitude} \cite{Kampf:2013vha}, which are the ``flavor-ordered'' amplitudes with flavor factors stripped away, much like the color-ordered partial amplitudes in QCD.  Partial amplitudes are symmetric under cyclic permutations of external particles. The 4-pt amplitude contains a single Feynman diagram involving the 4-pt interaction vertex as shown in Fig.~\ref{fig:4pt} and it is easy to see that the following amplitude satisfies the Adler's zero condition:
\be
\label{eq:4ptver}
M_4(p_1,p_2,p_3,p_4) = c \frac{s_{24}}{f^2} \to \frac{s_{24}}{f^2} \ ,
\ee
where we have used the notation $s_{ij}=(p_i+p_j)^2=2p_i\cdot p_j$ and absorbed the proportionality constant $c$ into the normalization of $f$. Upon momentum conservation, $M_4$ vanishes as any one of the external momenta is taken to zero.\footnote{An equivalent form commonly seen is $M_4=(s_{12}+s_{14})/f^2$.} One can then construct the 6-pt amplitudes by using the 4-pt interaction vertex given in Eq.~(\ref{eq:4ptver}). There are three contributions shown in Fig.~\ref{fig:6pt_part}, whose sum is
\be
\label{eq:6p_part}
\frac1{f^2}\left(\frac{s_{13} s_{46}}{P^2_{123}}+\frac{s_{24} s_{15}}{P^2_{234}}+\frac{s_{35} s_{26}}{P^2_{345}}\right) \ ,
\ee
where $P^2_{ijk}=(p_i+p_j+p_k)^2$. Obviously, the sum does not vanish as one external momentum is taken soft. The resolution is to introduce a 6-pt interaction vertex, shown in Fig.~\ref{fig:6pt_ver}, whose sole purpose is to satisfy the Adler's zero condition. If the following 6-pt vertex is added to  Eq.~(\ref{eq:6p_part}), as  shown in Fig.~\ref{fig:6pt_ver},
\be
\label{eq:6p_ver}
-\frac1{f^2}P^2_{135}\ ,
\ee
the resulting 6-pt amplitude 
\be
M_6= \frac1{f^2}\left(\frac{s_{13} s_{46}}{P^2_{123}}+\frac{s_{24} s_{15}}{P^2_{234}}+\frac{s_{35} s_{26}}{P^2_{345}}\right) -\frac1{f^2}P^2_{135} \ ,
\ee
vanishes as any one of the external momenta is taken to zero. Notice that Eq.~(\ref{eq:6p_ver}) is totally symmetric in cyclic permutations of  external particles, after imposing the momentum conservation. This ``soft bootstrapping'' was carried out up to 8-pt amplitudes in Ref.~\cite{Susskind:1970gf} and  completed to an arbitrary number of external legs in Refs.~\cite{Cheung:2014dqa,Cheung:2016drk}. In this approach, all interaction vertices are completely determined by starting with the 4-pt vertex in Eq.~(\ref{eq:4ptver}) and repeatedly requiring the Adler's zero condition for all higher point amplitudes. The only free parameter resides in the proportionality constant $c$ in Eq.~(\ref{eq:4ptver}), which was absorbed into the normalization of $f$. The most important lesson from this exercise, for the purpose of our discussion, is that the tree amplitudes and interaction vertices are constructed without ever referring to a coset $G/H$, as long as the notion of ``flavor ordering" exists.

How does one realize the above diagrammatic argument algebraically at the Lagrangian level? It turns out that the Adler's zero condition follows from the Ward identity of the shift symmetry  in Eq.~(\ref{eq:shift}) \cite{Low:2015ogb}. This can be understood intuitively as follows: the shift symmetry forbids non-derivative interactions in the Lagrangian so that interaction vertices in the Feynman diagrams carry positive powers of external momenta. Therefore, when one external momentum is taken to zero, the S-matrix element vanishes unless one of the internal propagator goes on-shell, which requires the presence of a cubic vertex \cite{Low:2015ogb}. However, for symmetric cosets employed in most, if not all, of the composite Higgs models, the internal $Z_2$ automorphism forbids cubic couplings among three Nambu-Goldstone bosons \cite{Low:2009di}. As a consequence, the shift symmetry in Eq.~(\ref{eq:shift}) implies the Adler's zero condition.

For a set of scalars transforming non-trivially under a linearly realized symmetry group $H$, Eq.~(\ref{eq:shift}) need to be expanded. More specifically, 
consider a set of scalars $\pi^a$ furnishing a linear representation of  $H$, $\pi^a \to \pi^a + i \alpha^i (T^i)_{ab} \pi^b + {\cal O}(\alpha^2)$, where $T^i$ is the generator of $H$. It is convenient to choose a basis where $T^i$ is purely imaginary and anti-symmetric, $(T^i)^T=-T^i$ and $(T^i)^{*}=-T^i$. Then at the next-to-leading order in $1/f^2$, the shift symmetry in Eq.~(\ref{eq:shift}) becomes \cite{Low:2014nga,Low:2014oga},
\be
\label{eq:shiftnlo}
\pi^a \to \pi^a +\varepsilon^a - \frac{c}{3f^2} (T^i)_{ab}(T^i)_{cd}\, \pi^b \pi^c \varepsilon^d \ ,
\ee
where $c$ is an arbitrary constant related to the normalization of $f$. Due to the anti-symmetricity of $T^i$, Eq.~(\ref{eq:shiftnlo}) has the property that the $1/f^2$ term vanishes if all but one Goldstone boson is set to zero. This ensures on-shell amplitudes of same-flavor Goldstone bosons satisfy both the Adler's zero condition and the Bose symmetry. The two-derivative Lagrangian satisfying the next-to-leading order shift symmetry is
\be
\label{eq:Lnlo}
{\cal L}= \frac12 \partial^\mu \pi^a \partial_\mu \pi^a + \frac{c}{6f^2} (T^i)_{ab}(T^i)_{cd}\ \partial^\mu \pi^a \partial_\mu \pi^c\ \pi^b \pi^d \ .
\ee
One can check that the 4-pt flavor-ordered amplitudes from Eq.~(\ref{eq:Lnlo}) gives precisely Eq.~(\ref{eq:4ptver}).

Going to higher orders, it will be convenient to define the matrix ${\cal T}$:
\be
{\cal T}_{ab} =\frac2{f^2} (T^i)_{ac}(T^i)_{db}\, \pi^c \pi^d \ ,\label{eq:mtdfg}
\ee
where we have chosen $c=2$ so as to conform with the convention in the literature in composite Higgs models. Then the nonlinear shift symmetry that enforces the Adler's zero condition to all orders in $1/f$ is \cite{Low:2017mlh,Low:2018acv}
\be
 \pi^{a\,\prime} = \pi^a + [F_1 (\mt)]_{ab}\  \vep^b\ ,\quad {F}_1({\cal T})= \sqrt{{\cal T}}\cot\sqrt{{\cal T}} \ , \label{eqshift}
\ee
and the two-derivative Lagrangian invariant under the nonlinear shift symmetry is
\be
 {\cal L}^{(2)} = \frac12 [F_2(\mt)^2]_{ab} \ \partial_\mu\pi^a \partial^\mu\pi^b\ , \quad  {F}_2({\cal T})=\frac{\sin\sqrt{{\cal T}}}{\sqrt{{\cal T}}} \label{eqnlsmlagep}\ .
 \ee
The 6-pt vertex in Fig.~\ref{fig:6pt_ver} arises from the $1/f^4$ term in ${\cal L}^{(2)}$, which is introduced so that the Lagrangian is invariant under the shift symmetry in Eq.~(\ref{eqshift}) up to the order of $1/f^4$. Therefore, one sees how the nonlinear shift symmetry implements the soft bootstrapping at the Lagrangian level.

The Lagrangian ${\cal L}^{(2)}$ is written entirely using generators of $H$, without reference to any coset $G/H$, as long as the linear representation furnished by $\pi^a$ satisfies \cite{Low:2014nga,Low:2014oga}
\be
\label{Eq:closure}
(T^i)_{ab}(T^i)_{cd}+(T^i)_{ac}(T^i)_{db}+(T^i)_{ad}(T^i)_{bc}=0\ .
\ee
This is a consistency condition imposed on the class of representations in which the low-energy effective Lagrangian can be constructed using only IR data. A direct comparison with the conventional CCWZ approach using a particular coset $G/H$ can be made upon the identification
\be 
(T^i)_{ab}= -if^{iab} = {\rm Tr}(T^i[X^a, X^b])\ .
\ee
Then it is clear that Eq.~(\ref{Eq:closure}) corresponds to the Jacobi identity of the structure constants $f^{iab}$ of  $G/H$.

\subsection{Effective Lagrangian Up to $p^4$ and All Orders in $1/f$}

The effective Lagrangian of Nambu-Goldstone bosons involves the systematic expansion in two parameters: the decay constant $f$ and the number of derivatives $\partial_\mu$. The Lagrangian in Eq.~(\ref{eqnlsmlagep}) contains two derivatives and resums terms to all orders in $1/f$. The overall coefficient of $1/4$ is determined by requiring a canonically normalized kinetic term, while the particular form of the function $F_2$ is fixed by the nonlinear shift symmetry in Eq.~(\ref{eqshift}), up to an overall rescaling  of $f$. To go beyond the two-derivative Lagrangian, it is necessary to introduce two objects with well-defined transformation properties under the nonlinear shift symmetry,
\bea
d_\mu &\to& U \ d_\mu \ U^{-1}\ , \\
 E_\mu^i T^i &\to& U (E_\mu^i T^i) U^{-1} - i U\, \partial_\mu (U^{-1})\ , \label{eq:Emurule}
 \eea
where  $U\in H$ and its explicit form is irrelevant for our discussion. Note that the Goldstone covariant derivative $d_\mu^a$ transforms in the same representation as $\pi^a$ under $H$, while the gauge connection $E_\mu^i$ sits in the adjoint representation of $H$. Both of them can be expressed in terms of IR data only \cite{Low:2014nga,Low:2014oga},
\bea
\label{eq:dmuexp}
d_\mu^a(\pi, \partial) &=& \frac{\sqrt{2}}{f}[ {F}_2({\cal T})]_{ab}\,\partial_\mu\pi^b \ ,  \\
\label{eq:emuexp}
{E}_\mu^i(\pi, \partial) &=& \frac2{f^2}  \partial_\mu \pi^a [{F}_4({\cal T})]_{ab}\, (T^i\pi)^b  \ ,
\eea
where ${F}_2(\cal T)$ is defined in Eq.~(\ref{eqnlsmlagep}) and
\be
  {F}_4({\cal T})=-\frac{2i}{\cal T}\sin^2\frac{\sqrt{\cal T}}2 \ .
\ee
In this notation the two-derivative Lagrangian is simply
\be
{\cal L}^{(2)} = \frac{f^2}4 d_\mu^\dagger d^\mu \ ,
\ee
which is manifestly invariant under the nonlinear shift symmetry.

At ${\cal O}(p^4)$, it is a well-known result from chiral perturbation theory that, without gauging any subgroup of $H$, there are three independent operators that can be constructed out of $d^a_\mu$ and $E_\mu^i$ \cite{Gasser:1984gg}. First let's define 
  the field strength tensor
\be
E_{\mu\nu}^i = \partial_\mu E^i_\nu - \partial_\nu E^i_\mu -  f^{ijk} E^i_\mu E^j_\nu\ .
\ee 
Then the three operators can be chosen to be
\be
O_1 = \left[ {\rm Tr}(d_\mu d^\mu ) \right]^2 \ , \quad O_2 =  \left[ {\rm Tr}( d_\mu d_\nu ) \right]^2 \ , \quad O_3 ={\rm Tr}(E_{\mu\nu} E^{\mu\nu}) \ .
\ee
The effective Lagrangian up to ${\cal O}(p^4)$ and to all orders in $1/f$ is then
\be
{\cal L}^{(4)}= {\cal L}^{(2)} +\sum_i c_i \, O_i  \ ,
\ee
where $c_i$ are incalculable coefficients encoding our ignorance of the UV physics. On the other hand, all coefficients in the $1/f$ expansion are completely fixed by the IR physics enforced by the nonlinear shift symmetry. Gauging a subgroup of $H$ will introduce additional operators. In what follows we will restrict our attention to the case relevant for a composite Higgs boson, where an $SU(2)\times U(1)$ inside the $SO(4)$ subgroup of $H$ is gauged.

\subsection{The Fundamental Representation of $SO(4)$}

As explained in Section \ref{sect:Intro}, a viable composite Higgs model must have the 125 GeV Higgs transforming as the fundamental representation of the $SO(4)$ subgroup of $H$. Using the explicit expression of group generators in Appendix \ref{appen:so5}, it is easy to see that they satisfy the following completeness relation:
\bea
T^i_{ab} T^i_{cd} = \frac{1}{2} (\delta_{ad} \delta_{bc} - \delta_{ac} \delta_{bd}),
\eea
so that Eq. (\ref{eq:mtdfg}) can be written as
\bea
\mT_{ab} = \frac{1}{f^2} (\delta^{ab} \pi^c \pi^c - \pi^a \pi^b).\label{eq:mtdf4}
\eea
Then  $F_1$, $F_2$ and $F_4$ simplify to
\bea
\left[ F_1( \mt) \right]_{ab} &=& \delta^{ab} \frac{|\pi|}{f} \cot \frac{|\pi|}{f} - \frac{\pi^a \pi^b}{f^2} \frac{f^2}{|\pi|^2} \left( \frac{|\pi|}{f} \cot \frac{|\pi|}{f} - 1\right) ,\\
\left[ F_2( \mt) \right]_{ab} &=& \delta^{ab}  \frac{f}{|\pi|}  \sin \frac{|\pi|}{f} - \frac{\pi^a \pi^b}{f^2} \frac{f^2}{|\pi|^2} \left[  \frac{f}{|\pi|}  \sin \frac{|\pi|}{f} - 1\right],\\
\left[ {F}_4({\cal T}) \right]_{ab} &=&-2i\left[\delta^{ab} \frac{f}{|\pi|}\sin^2\frac{|\pi|}{2f} -\frac{\pi^a \pi^b}{f^2} \frac{f^2}{|\pi|^2} \left(\frac{f}{|\pi|}\sin^2\frac{|\pi|}{2f} - 1 \right)\right] \ ,
\eea
where $|\pi|^2\equiv \pi^a\pi^a$. Also notice $F_1(0)=F_2(0)=1$. After gauging the $SU(2)_L\times U(1)_Y$ subgroup of $SO(4)$, the two-derivative Lagrangian in the unitary gauge looks particularly simple,
\begin{align}
\label{eq:L2unitary}
{\cal L}^{(2)} &= \frac{1}{2} \partial_\mu h \partial^\mu h+ \frac{g^2f^2}{4} \sin^2( \theta + h/f)    \left(W^+_\mu W^{- \mu} + \frac{1}{2\cos^2 \theta_W} Z_\mu Z^\mu \right)\ ,
\end{align}
where $h$ is the 125 GeV Higgs. In particular $\sin\theta \equiv v/f$, where $v=246$ GeV, is the misalignment angle between the $G/H$ breaking and the electroweak symmetry breaking. This can be seen from reading off the mass of the electroweak gauge boson in Eq.~(\ref{eq:L2unitary}):
\be
m_W =\frac{m_Z}{\cos\theta_W}=\frac12 g v = \frac12 g f \sin\theta\ .
\ee
Sometimes it is convenient to expand the effective Lagrangian in $h/v$,
\be
\label{eq:L2exp}
{\cal L}^{(2)} =\frac{1}{2} \partial_\mu h \partial^\mu h+ b_{nh} \left(\frac{h}{v}\right)^n \left( m_W^2  W^+_\mu W^{-\mu} + \frac{1}{2} m_Z^2 Z_\mu Z^{\mu} \right) \ .
\ee
For example, the first two coefficients are
\be
\label{eq:notest1}
b_{h}=2\sqrt{1-\xi} \ , \qquad b_{2h}=1-2\xi \ ,
\ee
where $\xi\equiv v^2/f^2$. Moreover, every single coefficient $b_{nh}$ is fixed by the nonlinear shift symmetry, up to the normalization of $f$.

One important observation here is that, at the two-derivative level, HVV and HHVV couplings have the same tensor structure as predicted in the SM. Furthermore, the strength of both couplings are reduced from the SM, which corresponds to $v/f\to 0$, for a real-valued $f$. This is the case for all compact coset $G/H$. For a non-compact coset, $f$ is purely imaginary and the strength is enhanced. At the four-derivative level, however, new tensor structures will appear.

The gauging the electroweak $SU(2)_L\times U(1)_Y$ will introduce additional building blocks for the effective Lagrangian at ${\cal O}(p^4)$ level. Formally we can choose to gauge the full $SO(4)$. In Eq.~(\ref{eq:dmuexp}) this amounts to replacing $\partial_\mu \to D_\mu=\partial_\mu +i A_\mu $ and 
 $d_\mu^a$ now becomes
\be
d_\mu^a(\pi,D)=  \frac{\sqrt{2}}{f}  [ {\cal F}_2({\cal T})]_{ab} (D_\mu \pi)^b \ , \quad D_\mu=\partial_\mu  + i A_\mu^i T^i \ ,
\ee
where $T^i$ is the generator of $SO(4)$. The gauging of $SO(4)$ breaks the nonlinear shift symmetry, similar to the gauging of $U(1)_{em}$ in chiral perturbation theory. The effect of such a breaking can then be captured by treating the $SO(4)$ gauge field as a ``spurion" in the $\mathbf{4}$ of $SO(4)$. Then one sees that the following object  transforms covariantly in the same way as $d_\mu^a(\pi, \partial)$,
\be
 \frac{\sqrt{2}i}{f}  [ {\cal F}_2({\cal T})]_{ab} (T^i \pi)^b\, A_\mu^i \ ,
 \ee
 which nonetheless is not invariant under local $SO(4)$ gauge transformations. But this is easy to fix, by replacing the $SO(4)$ gauge field with the corresponding field strength tensor,
\be
(f_{\mu\nu}^-)^a =  \frac{\sqrt{2}i}{f}  [ {\cal F}_2({\cal T})]_{ab} (T^i \pi)^b\, F_{\mu\nu}^i \ .
\ee
On the other hand, since the $SO(4)$ gauge field $A_\mu$ transforms under a local $H$-rotation in the same way as in Eq.~(\ref{eq:Emurule}), $E_\mu$ is now modified to be
\be
E_\mu^i (\pi, D) = A_\mu^i+ \frac{2}{f^2} D_\mu \pi^a \ [F_4({\cal T})]_{ab} (T^i\pi)^b \ .
\ee 
One can identify another spurion in the adjoint representation of $SO(4)$ and construct the following covariant object:
\be
(f_{\mu\nu}^+)^i = F_{\mu \nu}^i + \frac2{f^2}F_{\mu\nu}^j (T^j \pi)^a [ {\cal F}_4({\cal T})]_{ab}(T^i\pi)^{b} \ .
\ee
Both $f_{\mu\nu}^\pm$ transforms under local $H$-transformation as
\be
f_{\mu\nu}^\pm \to U\, f_{\mu\nu}^\pm\, U^{-1} \ .
\ee 
 Again both operators are constructed using only the infrared data, without recourse to a coset $G/H$. Here we follow the notation in Ref.~\cite{Contino:2011np}, which constructed the corresponding operators in the CCWZ formalism.

Given that $SO(4)\sim SU(2)_L\times SU(2)_R$, the adjoint representation of $SO(4)$ transforms as $(\mathbf{1},\mathbf{3})+(\mathbf{3},\mathbf{1})$ under $SU(2)_L\times SU(2)_R$. Then each object carrying the $SO(4)$ adjoint index can be divided into two categories, according to whether it transforms under $SU(2)_L$ or $SU(2)_R$. In addition, recall that an $SU(2)\times U(1)$ subgroup is identified with the electroweak gauge group. Our convention of $SO(4)$ and gauged generators are listed in Appendix \ref{appen:so5}. In the end the expressions for $d^a_\mu, E_\mu^i, (f_{\mu\nu}^-)^a$ and $(f_{\mu\nu}^+)^i$ for $SO(4)$  in the unitary gauge are given by
\bea
 d_\mu^{{a}} &= &\sqrt{2} \left[{\delta^{{a} 4}}\, \partial_\mu\left( \frac{h}f\right)
+ \frac{ \delta^{ar}}{2} \sin(\theta+h/f)  (W_\mu^r - \delta^{r3} B_\mu)\right]\ ,  \\
 (E_{\mu}^{L/R})^r &=& \frac{1 \pm \cos (\theta+h/f) }{2} W_\mu^r+ \frac{1\mp \cos(\theta+h/f)}{2} B_\mu \delta^{r3}\ , \\
(f_{\mu \nu}^-)^{{a}}&=&\frac1{\sqrt{2}}  \sin(\theta+h/f)  (W_{\mu \nu}^r - \delta^{r3} B_{\mu \nu}) \delta^{ra},\nonumber \\
(f_{\mu \nu}^{+L/R})^{r} &=& \frac{1\pm \cos (\theta+h/f)}{2} W^r_{\mu \nu} + \frac{1\mp \cos (\theta+h/f)}{2} \delta^{r3} B_{\mu \nu} \ ,
\eea
where the superscripts $L$ and $R$ refer to the upper and lower signs, respectively, and $r=1,2,3$ is the adjoint index in $SU(2)_{L/R}$.

Using these building blocks, one can construct 11 independent operators at ${\cal O}(p^4)$  \cite{Contino:2011np}, seven of which are even under space inversion $\vec{x}\to -\vec{x}$ and not contracted with $\epsilon_{\mu\nu\rho\sigma}$. We will focus on these seven CP-even operators in this work, compute them to all orders in $1/f$ and study their phenomenological consequences.  They can be written as
\begin{align}
\label{eq:op4}
O_1 & =  \left( d_\mu^{{a}}d^{\mu a} \right)^2\ ,\\
 O_2 &=  (d_\mu^{{a}} d_\nu^{{a}})^2\ ,  \\
O_3 &= \left[\left(E_{\mu \nu}^{L} \right)^r \right]^2- \left[\left(E_{\mu \nu}^{R} \right)^r \right]^2\ ,  \\
O_4^\pm &= -i\, d_\mu^a d_\nu^b \left[ (f^{+L}_{\mu \nu})^r\, T_L^{r} \pm (f^{+R}_{\mu \nu})^r\, T_R^{r} \right]_{ab}\ ,  \\
O_5^+&=\left[(f_{\mu \nu}^-)^{{a}}\right]^2  \ , \\
 O_5^-&=\left[(f_{\mu \nu}^{+L})^{r}\right]^2-\left[(f_{\mu \nu}^{+R})^{r}\right]^2 ,
\end{align}
where $T_{L/R}^r$ is the $SU(2)_{L/R}$ generator Appendix \ref{appen:so5}. 

The power counting of the four-derivative Lagrangian is governed by the naive dimensional analysis (NDA) \cite{Manohar:1983md}, which states that each Nambu-Goldstone field $\pi$ is suppressed by $f$, while the (gauge covariant) derivative $D_\mu$ is suppressed by $\Lambda\sim 4\pi f$,
\be
\label{eq:NDAcount}
S^{(4)} = \int d^4x \Lambda^2 f^2\ {\cal L}^{(4)}\left(\frac{\pi}{f}, \frac{D}{\Lambda}\right) = \int d^4x \sum_i \frac{c_i}{16\pi^2} O_i \ ,
\ee 
where $c_i$ are expected to be order unity constants parameterizing the incalculable UV physics at the scale $\Lambda \sim 4\pi f$. In some cases operators  contributing to  couplings of neutral particles and the on-shell photon  are further suppressed by additional loop factors. 

In composite Higgs models there are typically additional resonances at the scale $m_\rho = g_\rho f$, where $1 \lesssim g_\rho \lesssim 4\pi$ characterizes the coupling strength associated with the strong dynamics. After integrating out these resonances, the nonlinear structure of the effective Lagrangian remains the same, with the $\Lambda$ scale suppressing the derivative operator $D_\mu$ replaced by $m_\rho$ \cite{Giudice:2007fh}:
\be
\label{eq:lag4}
S_{SILH}^{(4)}= \int d^4x \ m_\rho^2\, f^2\ {\cal L}^{(4)}\left(\frac{\pi}f, \frac{D}{m_\rho}\right)=\int d^4x\ \sum_i \frac{c_i}{g_\rho^2} O_i \ ,
\ee 
Notice that when $g_\rho$ saturates $4\pi$, the 
``SILH" power counting reverts to NDA in Eq.~(\ref{eq:NDAcount}). There are certainly additional effects that break the nonlinear shift symmetry, such as the Higgs potential and the Higgs coupling to fermions. However, as argued in Ref.~\cite{Liu:2018vel}, they would modify the nonlinearity in Eq.~(\ref{eq:lag4}) only at the loop-level.

\noindent
\section{Universal Relations}
\label{sect:3}

The effective Lagrangian in Eq.~(\ref{eq:lag4}) describes the nonlinear interaction of a composite Higgs boson with the electroweak gauge bosons to all orders in $1/f$ and up to the four-derivative level. It is most convenient to express the Lagrangian in unitary gauge, 
\beq
\label{eq:Cis}
\mL_{\rm NL} =  \sum_i\frac{m_W^2}{m_\rho^2} \left(C^h_i \mI^h_i  +C^{2h}_i \mI^{2h}_i +  C^{3V}_i \mI^{3V}_i  \right)\ ,
\eeq
where $\mI_i^{h}$ and $\mI_i^{2h}$ are ${\cal O}(p^4)$ operators contributing to HVV and HHVV couplings, respectively, while $\mI^{3V}_i$  contains triple gauge boson couplings (TGC) up to one Higgs boson. We use the notation $V=W,Z,\gamma$. Note that we have factorized out a normalization factor $m_W^2/m_\rho^2$ in the coupling coefficients $C_i$, which are linear combinations of the dimensionless coefficients $c_i$ in \Eq{eq:lag4}.  For the anomalous triple gauge boson couplings (aTGC), we  added tilde to the notation used in Ref.~\cite{Hagiwara:1986vm},  so as to emphasize our choice of normalization factor  $m_W^2/m_\rho^2$ in Eq.~(\ref{eq:Cis}). The two conventions are related by
\beq
\label{eq:TGCdef}
\delta g_1^Z = \frac{m_W^2}{m_\rho^2} \delta \tilde g_1^Z, \qquad \delta \kappa_\gamma = \frac{m_W^2}{m_\rho^2} \delta \tilde \kappa_\gamma, \qquad \delta \kappa_Z = \frac{m_W^2}{m_\rho^2} \delta \tilde \kappa_Z\ .
\eeq
More specifically, operators in $\mI_i^{h}$ involve the following structures
\be
\label{eq:lstruc1}
\frac{h}{v} V_{1\,\mu} \ewd^{\mu\nu} V_{2\,\nu} \ ,\qquad \frac{h}{v} V_{1\,\mu\nu}  V_2^{\mu\nu} \ ,
\ee
where  $\ewd^{\mu\nu} = \partial^\mu\partial^\nu-\eta^{\mu\nu}\partial^2$ and $V_{1/2} \in\{W, Z,\gamma\}$. Those in  $\mI_i^{2h}$ are given by
\be
\frac{h^2}{v^2} V_{1\,\mu} \ewd^{\mu\nu} V_{2\,\nu}\ , \quad  \frac{h^2}{v^2} V_{1\,\mu\nu}  V_2^{ \mu\nu} \ , \qquad \frac{\partial_\mu h \partial_\nu h}{v^2} V_1^\mu V_2^{\nu} \ .
\label{eq:lstruc2}
\ee
Lastly, operators in the $\mI_i^{3V}$ class are of the form
\be
\label{eq:lstruc3}
V_{1\,\mu} V_{2\,\nu} V_3^{\mu\nu} \ , \qquad \frac{h}{v} V_{1\,\mu} V_{2\,\nu} V_3^{ \mu\nu} \ , \qquad \frac{\partial_\mu h}{v} V_{1\,\nu} V_2^{\mu} V_3^{\nu} \ .
\ee
The complete list of operators in each category is listed in Tables~\ref{tab:oneh}-\ref{tab:tgc}, where we have computed the coupling $C_i$ in terms of the $c_i$ coefficient in  \Eq{eq:lag4}. There are 6 operators in $\mI_i^h$, 10 in $\mI_i^{2h}$ and 9 in $\mI_i^{3V}$. These three tables summarize the predictions of universal nonlinearity (NL) from a composite Higgs in Higgs couplings with electroweak gauge bosons.

It is instructive to compare with modifications in HVV and HHVV couplings from  the Standard Model Effective Field Theory (SMEFT) \cite{Henning:2014wua}, which augments the SM Lagrangian with higher dimensional operators with arbitrary coefficients. In particular, the dim-6 and dim-8 operators relevant for our discussion are parameterized as follows: 
\be
\label{eq:smeft}
{\cal L}_{\text{SMEFT}} \supset \sum_{i=W,B,HW,HB} \frac{c_{i}}{m_\rho^2}\,  \mO_{i} +   \frac{c_{i}^8}{f^2 m_\rho^2} \, (H^\dagger H)\mO_{i}\ ,
\ee
where $\mO_W,\mO_B,\mO_{HW},\mO_{HB}$ are defined explicitly in \Eq{eq:d6ops}. The corresponding contributions to $\mI_i^{h/2h/3V}$ from dim-6 operators (D6) in SMEFT are also given in Tables \ref{tab:oneh}--\ref{tab:tgc}. The matching of the $\mO(p^4)$ operators in \Eq{eq:op4} to  SMEFT at the dimension-6 order is done explicitly in  Appendix \ref{app:d6}.

\begin{table}[!t]
\begin{center}
\begin{tabular}{|l|c |c| c| c| c|c|c|}
\hline
$\mI^{h}_i$ &   $  C^h_i$ (NL)& $  C^h_i$ (D6)\\
 \hline
(1) $  \frac{h}{v} Z_{\mu} \ewd^{\mu \nu} Z_{\nu}$ & $ \begin{array}{c}\frac{4 c_{ 2 w} }{c^2_{w}}  \left(-2 c_3  + c_4^- \right)  \\
+\frac{4}{c^2_{w}}   c_4^+\cos \theta \end{array}$ &$ \begin{array}{c}2 (c_W + c_{HW}) \\+2 t^2 _{w} (c_B + c_{HB}) \end{array}$\\
 \hline
(2)  $ \frac{h}{v}  Z_{\mu\nu} Z^{\mu\nu}$ & $ \begin{array}{c} -\frac{2  c_{2 w}}{c^2_{w}}  \left( c_4^-  + 2 c_5^- \right)\\ -\frac{2 }{c^2_{w}}   \left( c_4^+  - 2 c_5^+ \right) \cos \theta   \end{array} $&$-( c_{HW} + t^2_{w} c_{HB})$\\
 \hline
(3)  $ \frac{h}{v}  Z_{\mu} \ewd^{\mu \nu} A_{\nu}$ & $8  \left( - 2 c_3  +   c_4^- \right)  t_{ w} $& $\begin{array}{c} 2 t_w (c_W + c_{HW})\\ - 2 t_w (c_B + c_{HB})  \end{array}$ \\
\hline
(4) $ \frac{h}{v}  Z_{\mu\nu} A^{\mu\nu}$ & $-4  \left(  c_4^-   +2 c_5^- \right)  t_{ w} $&  $ -  t_w(c_{HW}- c_{HB})$\\
\hline
(5)  $ \frac{h}{v}  W^+_{\mu} \ewd^{\mu \nu} W^-_{\nu} + h.c.$ & $\begin{array}{c}4 (-2 c_3  + c_4^- ) \\
+  4 c_4^+ \cos \theta \end{array}$ & $2 (c_W + c_{HW})$\\
 \hline
(6) $ \frac{h}{v}  W^+_{\mu\nu} W^{-\mu\nu} $ & $\begin{array}{c} -4( c_4^- + 2 c_5^-)  \\ -4 \left( c_4^+  - 2 c_5^+\right) \cos \theta \end{array}$&   $-2 c_{HW} $ \\
\hline
\end{tabular}
\end{center}
\caption{Single Higgs coupling coefficients $C_i^h$ for the non-linearity case (NL) and the  purely dimension-6 contributions (D6) in SMEFT. Here  $c_w, t_w$ denote $\cos\theta_W, \tan\theta_W$ respectively, where $\theta_W$ is the weak mixing angle. $\mD^{\mu\nu}$ denotes $\partial^\mu \partial^\nu - \eta^{\mu\nu} \partial^2$.}
\label{tab:oneh}
\end{table}

\begin{table}[t]
\begin{center}
\begin{tabular}{|l|c |c| c| c| c}
\hline
$\mI^{ 2 h}_i$ &  $ C^{ 2h}_i$ (NL) &  $   C^{2h}_i$ (D6) \\
 \hline
(1) $  \frac{h^2}{v^2} Z_{\mu} \ewd^{\mu \nu} Z_{\nu}$ & $ \begin{array}{c} \frac{2c_{ 2 w} }{c^2_{w}} \left( - 2 c_3  +  c_4^-  \right) \cos \theta   \\ +   \frac{2 }{c^2_{w}} c_4^+ \cos 2 \theta  \end{array}$ & $\frac12 C_1^h$ 
 \\
 \hline
(2) $  \frac{h^2}{v^2} Z_{\mu\nu} Z^{\mu\nu}$ & $\begin{array}{c} - \frac{c_{2 w}}{c^2_{w}}  \left(c_4^-  + 2 c_5^-  \right) \cos \theta  \\ - \frac{1}{c^2_{w}}  \left(c_4^+ - 2 c_5^+ \right) \cos 2 \theta  \\ \end{array}  $ &$\frac12 C_2^h$ \\
 \hline
(3) $  \frac{h^2}{v^2} Z_{\mu} \ewd^{\mu \nu} A_{\nu}$ & $ 4  t_{ w}  \left( -2c_3   + c_4^- \right) \cos \theta $ & $\frac12 C_3^h$ \\
 \hline
(4) $  \frac{h^2}{v^2} Z_{\mu\nu} A^{\mu\nu}$ & $ -2  t_{ w}   \left(   c_4^-  + 2 c_5^-  \right)  \cos \theta $ &   $\frac12 C_4^h$  \\
    \hline
(5)  $ \frac{h^2}{v^2} W^+_{\mu} \ewd^{\mu \nu} W^-_{\nu} + h.c.$ & $ \begin{array}{c} 2  (- 2c_3 + c_4^- ) \cos \theta \\ + 2 c_4^+ \cos 2 \theta   \end{array}$ & $\frac12 C_5^h$ \\
 \hline
(6) $ \frac{h^2}{v^2} W^+_{\mu\nu} W^{-\mu\nu}$ & $ \begin{array}{c} -2 \left( c_4^- + 2 c_5^- \right)\cos \theta  \\ -2  \left(c_4^+  - 2 c_5^+ \right) \cos 2 \theta  \end{array}$ &  $\frac12 C_6^h$\\
\hline
(7) $      \frac{(\partial_\nu h)^2} {v^2}Z_\mu Z^{ \mu}  $& $\frac{8 }{c^2_{w}}  c_1 \sin^2\theta$ &  $\times$\\
\hline
(8)  $   \frac{\partial_\mu h \partial_\nu h}{v^2} \, Z^\mu Z^\nu  $& $\frac{8 }{c^2_{w}}  c_2  \sin^2\theta$&  $\times$\\
\hline
(9) $      \frac{(\partial_\nu h)^2}{v^2} W^+_\mu W^{- \mu}  $& $16 c_1 \sin^2\theta$ & $\times$ \\
\hline
(10) $  \frac{ \partial^\mu h \partial^\nu h }{v^2}\, W^+_{\mu} W^{-}_\nu  $& $16  c_2  \sin^2\theta$ &  $\times$\\
\hline
\end{tabular}
\end{center}
\caption{The coupling coefficients $C_i^{2h}$ involve two Higgs bosons  for universal  nonlinearity case (NL) and the  dimension-six case in SMEFT (D6). A cross ($\times$) means there is no contribution at the order we considered. Notice $C_i^{2h}=C_i^h/2$ for  SMEFT at the dimension-6 level.}
\label{tab:twoh}
\end{table}

\begin{table}[t]
\begin{center}
\begin{tabular}{|l|c |c| c| c| c}
\hline
$\mI^{3V}_i$ & $  C_i^{3V}$ (NL)& $  C_i^{3V}$ (D6) \\
\hline
$(\delta \tilde{g}_1^Z) \begin{array}{c} i g c_w W^{+\mu \nu}  W^-_\mu  Z_\nu  \\
+ h.c. \end{array}$ &  $ \begin{array}{c}- \frac{2}{c^2_{w}} \left[  \left(- 2c_3 + c_4^- \right) \cos \theta + c_4^+  \right]   \end{array}$  & $- \frac{c_W + c_{HW}}{c^2_{w}}$ \\
\hline
$(\delta \tilde \kappa_\gamma)i e W^{+}_{\mu} W^{-}_{ \nu} A^{\mu \nu}  $ &   $ - 4\left(  c_4^+  - 2  c_5^+ \right)    $ &  $  -\left( c_{HW} + c_{HB}\right)$ \\
\hline
$(\delta \tilde \kappa_Z)i g  c_wW^{+}_{\mu} W^{-}_{ \nu}   Z^{\mu \nu} $& $ \begin{array}{c}- \frac{2}{c^2_{w}} \left(-2 c_3 + c_4^- \right) \cos \theta  \\ - \frac{2}{c^2_{w}}(  c_4^+ c_{2 w}   + 4 c_5^+ s^2_{w} ) \end{array}  $ & $- \frac{c_W}{c^2_{w}}-c_{HW} + t^2_{w} c_{HB}$\\
\hline
$(1) \begin{array}{c} i g c_w \frac{h}{v}W^{+\mu \nu} W^-_\mu  Z_\nu\\
 + h.c. \end{array}$& $ \begin{array}{c}
-\frac{4}{c^2_w} \left[(-2 c_3 + c_4^-) (1-\frac 32 \sin^2\theta) \right.\\
\quad\left.  + c_4^+ \cos \theta \right] \\
+16  ( c_3 + c_5^- - c_5^+ \cos \theta)  
\end{array}$   &  $-\frac{2}{c_w^2}(c_W + c_{HW}) - 4 c_W$ \\
\hline
$(2)\begin{array}{c}i  e\frac{h}{v} W^{+\mu \nu} W^-_\mu  A_\nu\\
 + h.c.\end{array}$& $ 16 (c_3 + c_5^- - c_5^+ \cos \theta)  $  & $-4 c_W$\\
\hline
$(3)i g c_w \frac{h}{v}W^{+}_{\mu} W^{-}_{ \nu}  Z^{\mu \nu} $& $ \begin{array}{c}
- \frac{4}{c_w^2}\left(1-\frac32 \sin^2\theta\right)  (-2c_3 + c_4^-)  \\
 +  16(c_3 + c_5^-)   \\
- \frac{4}{c_w^2} ( c_4^+ c_{2 w} +4 c_5^+  ) \cos \theta \\
\end{array}$  &  $\begin{array}{c}-\frac{2(1 + 2 c_w^2)}{c_w^2}c_W- 2 c_{HW} \\
+ 2  t_w^2 c_{HB}\end{array}$ \\
\hline
$(4)i e\frac{h}{v}W^{+}_{\mu} W^{-}_{ \nu} A^{\mu \nu}  $& $16 (c_3 + c_5^-) - 8c_4^+ \cos \theta  $ & $-4 c_W - 2 c_{HW} - 2 c_{HB}$ \\
\hline
$\begin{array}{lc}(5)i  eW^+_{[\mu} W^-_{\nu]} A^\mu \frac{\partial^\nu h}{v}\\
(6)-ig^{\prime} s_w W^+_{[\mu} W^-_{\nu]} Z^\mu \frac{\partial^\nu h}{v}
\end{array}$& $  \begin{array}{c}-8 \left[(-2 c_3  + c_4^-) + \cos\theta c_4^+\right.\\
\left.\quad -  \sin^2\theta c_3  \right]\end{array}$   & $ -4\left(c_W + c_{HW}\right) $ \\
\hline
\end{tabular}
\end{center}
\caption{Triple gauge boson couplings involving one or no Higgs.}
\label{tab:tgc}
\end{table}

The most important observation for our purpose is that there are only 6 unknown  $c_i$'s in Eq.~(\ref{eq:lag4}), which parameterize effects of the incalculable  ultraviolet physics. These 6 coefficients enter into Tables~\ref{tab:oneh}-\ref{tab:tgc}, which contain a total of 25 operators entering HVV, HHVV, HVVV and TGC couplings. In principle, these operators can all be measured experimentally through angular distributions of decay products, with varying degrees of precision. ``Universal relations" are precisely relations among the coefficients of the 25 operators listed in Tables~\ref{tab:oneh}-\ref{tab:tgc} that are independent of the unknown $c_i$ coefficients. They depend on only one  parameter $\sin\theta=v/f$, or equivalently the normalization of the decay constant $f$, and are insensitive to the coset structure $G/H$ invoked in the UV.

From Tables \ref{tab:oneh}--\ref{tab:tgc} one can see that, for SMEFT at the dimension-6 level, $C_i^{h}$ and $C_i^{2h}$ are related
\be
C_i^{2h} = \frac12 C_i^h \quad \, i=1, \cdots, 6 \ .
\ee
This is because at this order the operators involve to the combination $H^\dagger H \sim (h+v)^2 = h^2+ 2 v h + v^2$, which gives the relation above. It will not hold anymore at the dimension-8 or higher, which involves a higher power in $H^\dagger H$. In addition, the term involving $v^2$ does not give rise to any new relations among the Wilson coefficients because this term usually goes into re-defining the SM couplings that are used as input parameter experimentally. We demonstrate this subtlety explicitly in the case of the HVVV couplings in Appendix \ref{app:d6}. Furthermore, it is interesting to observe in Tables \ref{tab:oneh} and \ref{tab:twoh} the following relations involving HZZ and HHZZ couplings,
\bea
C_1^{2h}&=&\frac12 C_1^h+ \frac{2}{c_w^2}c_4^+\left(\cos2\theta - \cos\theta\right) \ ,\\
C_2^{2h}&=&\frac12 C_1^h+ \frac{1}{c_w^2}(c_4^+-2c_5^-)\left(\cos2\theta - \cos\theta\right)\ ,
\eea
which involve non-calculable $c_i$ coefficients in Eq.~(\ref{eq:lag4}). Phenomenologically this has an important implication: the deviation in HZZ coupling could be accidentally small, while the HHZZ correction could still be sizeable. Similar considerations apply to HWW and HHWW couplings as well. On the other hand, the HZ$\gamma$ coupling is strongly correlated with HHZ$\gamma$ coupling.

Below we present some examples of universal relations in composite Higgs models.
In this regard, notice that the photon couplings  are not modified by the vacuum misalignment angle $\theta$ due to the unbroken electromagnetic gauge invariance. Therefore their coefficients can be related directly to the Wilson coefficients $c_i$ in Eq.~(\ref{eq:lag4}) without prior knowledge of $\sin\theta=v/f$,
\beq
\label{eq:relation}
 C_3^h= 8t_{w} (-2 c_3 + c_4^-) , \qquad  C_4^h=4t_w(c_4^- + 2 c_5^- ), \qquad \delta \tilde \kappa_\gamma=-4(c_4^+ - 2 c_5^+)  \ .
\eeq
Using these relations, it is straightforward to derive the following three universal rations involving only the HVV and aTGC couplings:
\bea
\label{eq:ident1}
{\rm UR1}&:& \frac{ C^{h}_6 - C^{h}_4/t_{w} }{\delta \tilde\kappa_\gamma}= \frac{ 2c^2_{w}\,  C^{h}_2 -c_{2w} \,C^{h}_4/t_{w} }{\delta \tilde\kappa_\gamma}= \cos\theta \approx 1-\frac12 \xi , \\
\label{eq:ident2}
{\rm UR2}&:&\frac{c_{2w}}{t_w}C^{h}_3 - 2  c_w^2 \,C^{h}_1  = 4 c^2_w\, \delta \tilde{g}_1^Z \, \cos\theta + \frac{1}{t_w}C^{h}_3\, \cos^2\theta \ , \\
{\rm UR3}&:& C_5^h  = -2 c^2_w\, \delta \tilde{g}_1^Z\,  \cos\theta + \frac{1}{2t_w}C^{h}_3\, \sin^2\theta \ ,
\eea
where $\xi =v^2/f^2= \sin^2\theta$.

The second class  of universal relations involve HVV and HHVV couplings. These relations come about naturally in composite Higgs models because of the nonlinear symmetry (or broken symmetry in the CCWZ formalism) relating one to the other.
In this case we found the following useful relations  as signs of universal nonlinearity,
\bea
\label{eq:ident3}
{\rm UR4}&:& \frac{C^{2h}_3}{C^h_3} = \frac{C^{2h}_4}{C^h_4} =\frac12 \cos\theta  \ ,\\
\label{eq:ident4}
{\rm UR5}&:&\frac{C^{2h}_5 - C^{2h}_3 /2t_w} {C^h_5 - C^{h}_3/2t_{w}} =  \frac{C^{2h}_6 -C^{2h}_4/t_{w} }{C^h_6 - C^{h}_4/t_{w}} = \frac{\cos2\theta}{2\cos\theta} \approx \frac12\left(1 - \frac32 \xi \right)\, ,\\
{\rm UR6}&:&\frac{s_{2w} \,C^{2h}_1 - c_{2w}\,   C^{2h}_3} {s_{2w}\, C^h_1-c_{2w} \,  C^{h}_3} =\frac{s_{2w}\, C^{2h}_2 - c_{2w}\,   C^{2h}_4} {s_{2w}\, C^h_2-c_{2w}\,   C^{h}_4}  = \frac{\cos2\theta}{2\cos\theta}\approx \frac12\left(1 - \frac32 \xi \right)\ .
\eea
In addition there is one universal relation involving HVV and HVVV couplings:
\beq
{\rm UR7}: \frac{(C^{h}_3 - 2 t_w C^{h}_5 )- s_{2w} (C_1^{3V} - C_2^{3V})} {C_3^h}  = 1 - \frac32 \sin^2 \theta\ .
\eeq
As emphasized already, these relations are all determined by the one single input parameter $\sin\theta$ and free from the incalculable coefficients in Eq.~(\ref{eq:lag4}).

One can compare the prediction of universal nonlinearity with that from SMEFT with arbitrary Wilson coefficients. For the purpose of demonstration, we consider UR4 and UR6. Using the parameterization in Eq.~(\ref{eq:smeft}), they become at the leading order in $\xi$,
\bea
\label{eq:ident}
&&\frac{C^{2h}_3}{C^h_3}\approx  \frac12 \left(1 + \frac{c_{HW}^8 - c^8_{HB}}{c_{HW} - c_{HB}}\xi \right),\\
&& \frac{C^{2h}_4}{C^h_4} \approx \frac12 \left(1 + \frac{c_{W}^8 - c^8_{B} + c_{HW}^8 - c^8_{HB} }{c_W - c_B + c_{HW} - c_{HB}}\xi\right),\\
&&\frac{s_{2w}C^{2h}_1 - c_{2w}   C^{2h}_3} {s_{2w}C^h_1-c_{2w}   C^{h}_3} \approx\frac12 \left(1 + \frac{c_{W}^8 + c^8_{B} + c_{HW}^8 + c^8_{HB} }{c_W + c_B + c_{HW} +c_{HB}}\xi\right),\\
&&\frac{s_{2w} C^{2h}_2 - c_{2w}   C^{2h}_4} {s_{2w} C^h_2-c_{2w}   C^{h}_4}  \approx \frac12 \left(1 + \frac{ c_{HW}^8 + c^8_{HB} }{c_{HW} + c_{HB}}\xi\right).
\eea
These relations make it clear that predictions of universal nonlinearity start appearing at the level of dimension-8 operators. More specifically, the nonlinear shift symmetry of a composite Higgs boson relates Wilson coefficients of certain dimension-8 operators with those of dimension-6 operators so that their ratios are fixed. In fact, at the two-derivative level, all Wilson coefficients in the $1/f$ expansion are uniquely determined, up to the normalization of $f$. This implies effects from purely dim-6 operators can be absorbed into a re-scaling of $f$. To make a statement on the universal nonlinearity it is necessary to include effects from dimension-8 operators.

\section{Testing the Universal Relations: A Preliminary Study}
\label{sect:4}

In order to test the universal relations experimentally, it is necessary to connect the ${\cal O}(p^4)$ operators in Tables~\ref{tab:oneh}-\ref{tab:tgc} with observables. There are four classes of couplings that participate in the universal relations: HVV, TGC, HHVV and HVVV, and we need to measure them up to ${\cal O}(p^4)$ with great precision, as the effect of nonlinearity only shows up at the level of dim-8 operators. (For recent studies on effects of some class of dim-8 operators, see Refs.~\cite{Hays:2018zze,Bellazzini:2018paj}.) The necessity of high precision makes it desirable to introduce new analysis techniques \cite{Brehmer:2016nyr,Guest:2018yhq}, which is beyond the scope of the present work.

\begin{table}[t]
\centering
\begin{tabular}{|c|c |}
\hline
 Operator &  Tensor Structure\\
 \hline
$ \begin{array}{cc}
h Z_{\mu} \ewd^{\mu \nu} Z_{\nu} \\
hW^+_{\mu} \ewd^{\mu \nu} W^-_{\nu}+ h.c. \\
\end{array} $& $i\left[( p_1^2 + p_2^2) g^{\mu\nu}-p_1^\mu p_1^\nu - p_2^\mu p_2^\nu \right]$ \\
\hline
$h Z_{\mu} \ewd^{\mu \nu} A_{\nu} $ & $ i (p_2^2 g^{\mu\nu}-p_2^\mu p_2^\nu) $  \\
 \hline
$ \begin{array}{cc}
\frac12 h Z_{\mu\nu} Z^{\mu\nu}\\
h Z_{\mu\nu} A^{\mu\nu}\\
h W^+_{\mu\nu} W^{-\mu\nu}\\
\end{array}$ & $2 i (p_2^\mu  p_1^\nu-  g^{\mu\nu} p_1 \cdot p_2 )$\\
 \hline
\end{tabular}
\caption{The tensor structure of $h(p_3)V_1(p_1,\mu)V_2(p_2,\nu)$ operators. All the momenta are taken to be ingoing.\\  \label{tab:frhvv}}

\begin{tabular}{|c|c |}
\hline
 Operator &  Tensor Structure\\
 \hline
$\begin{array}{cc} i   W^{+\mu \nu} W^-_\mu Z_\nu + \hc \\
i   W^{+\mu \nu} W^-_\mu A_\nu + \hc
\end{array}$ & $ g^{\mu \nu}  (p_1 - p_2)^{ \rho} - g^{\mu \rho}  p_{1}^{ \nu}+ g^{\nu \rho}   p_{2}^{ \mu} $ \\
\hline
$\begin{array}{cc}i W^{+\mu} W^{- \nu}  Z_{\mu \nu} \\
i   W^{+\mu} W^{- \nu} A_{\mu \nu} 
 \end{array}$& $  p_3^\mu g^{\rho\nu}  -p_3^\nu g^{\mu\rho} $\\
  \hline \hline
 $\begin{array}{cc} i h  W^{+\mu \nu} W^-_\mu Z_\nu + \hc \\
i   hW^{+\mu \nu} W^-_\mu A_\nu + \hc
\end{array}$ & $ g^{\mu \nu}  (p_1 - p_2)^{ \rho} - g^{\mu \rho}  p_{1}^{ \nu}+ g^{\nu \rho}   p_{2}^{ \mu} $ \\
\hline 
$\begin{array}{cc}i hW^{+\mu} W^{- \nu} Z_{\mu \nu}  \\
i  hW^{+\mu} W^{- \nu} A_{\mu \nu}  
 \end{array}$& $   p_3^\mu g^{\rho\nu} -p_3^\nu g^{\mu\rho}$\\
\hline
$\begin{array}{cc} i W^+_{ \mu} W^-_{\nu } Z^\mu \partial^\nu h+h.c.\\
 i W^+_{ \mu} W^-_{\nu } A^\mu \partial^\nu h + h.c.\\
\end{array}$& $ p_4^\nu g^{\mu \rho} -p_4^\mu g^{\nu \rho} $\\
\hline
\end{tabular}
\caption{The tensor structure of $W^+(p_1,\mu)W^-(p_2,\nu)V(p_3,\rho)$ operators. The tensor structure of $h(p_4)W^+(p_1,\mu)W^-(p_2,\nu)V(p_3,\rho)$ operators.\\\label{tab:frvvv}}
\begin{tabular}{|c|c |}
\hline
 Operator &  Tensor Structure\\
 \hline
$ \begin{array}{cc}
h^2 Z_{\mu} \ewd^{\mu \nu} Z_{\nu} \\
h^2W^+_{\mu} \ewd^{\mu \nu} W^-_{\nu}+ h.c. \\
\end{array} $&$2 i[g^{\mu\nu} ( p_1^2 + p_2^2)- p_1^\mu  p_1^\nu - p_2^\mu p_2^\nu $]\\
\hline
$h^2 Z_{\mu} \ewd^{\mu \nu} A_{\nu}$ &$2 i[g^{\mu\nu}  p_2^2  - p_2^\mu p_2^\nu $]\\
 \hline
$ \begin{array}{cc}
h^2 W^+_{\mu\nu} W^{-\mu\nu}\\
\frac12 h^2 Z_{\mu\nu} Z^{\mu\nu}\\
h^2 Z_{\mu\nu} A^{\mu\nu}\\
\end{array}$ & $4i( p_2^\mu p_1^\nu-  g^{\mu\nu} p_1 \cdot p_2 )$\\
 \hline
 $  \begin{array}{cc}   
 \frac12 (\partial_\nu h)^2 Z_\mu Z^{ \mu}\\
  (\partial_\nu h)^2 W^+_\mu W^{- \mu} \\
  \end{array}$& $ -2 ig^{\mu\nu} p_3 \cdot p_4$ \\
\hline
$ \begin{array}{cc}  
\frac12\partial_\mu h \partial_\nu h \, Z^{\mu} Z^{\nu}  \\
\partial_\mu h \partial_\nu h \, W^{+\mu} W^{- \nu}  \\
\end{array}$& $- i(p_3^\mu  p_4^\nu  +  p_3^\nu p_4^\mu ) $\\
 \hline
\end{tabular}

\caption{The tensor structure of $h(p_3)h(p_4)V_1(p_1,\mu)V_2(p_2,\nu)$ operators.\\\label{tab:frhhvv}}
\end{table}

The first class, HVV couplings, received much of the attention and was the top priority at the LHC Run 1. The different operators can be probed by studying kinematic distributions in the decay product of single Higgs production \cite{Gainer:2011xz,Bolognesi:2012mm,Stolarski:2012ps,Artoisenet:2013puc,Gainer:2014hha}. Not surprisingly, most projections on  extracting Higgs couplings in future colliders also  focus on this class \cite{Craig:2015wwr,Durieux:2017rsg,Gu:2017ckc}. The second class, the TGC couplings, has also been studied extensively \cite{Hagiwara:1986vm,Degrande:2012wf,Falkowski:2016cxu}. The HHVV and HVVV couplings, to the contrary, seems to have escaped much of the attention. In particular, the HHVV coupling sits among the least tested sectors of the SM Higgs boson. A number of theoretical studies exist in the literature \cite{Dolan:2013rja,Dolan:2015zja,Bishara:2016kjn,Arganda:2018ftn,Kilian:2018bhs}, although none specifically addresses the issue of measuring the tensor structures, which requires including the complete list of operators up to ${\cal O}(p^4)$.
Therefore, it is clear that testing the universal relation is a long-term program and should be among the priorities of the future experimental program on the Higgs boson. In the remainder of this work, we will be content with a very preliminary phenomenological study on universal relations in composite Higgs.

As a first step toward studying phenomenological consequences of universal relations, in Tables~\ref{tab:frhvv}-\ref{tab:frhhvv} we give the Feynman rules for the interaction vertices listed in Tables~\ref{tab:oneh}-\ref{tab:tgc}. 
In the tables, we have taken all the momenta ingoing to the vertices. The Lorentz structures of the operators have already been spelled out in Eqs.~(\ref{eq:lstruc1})--(\ref{eq:lstruc3}). For HVV couplings, the structure $V_1^\mu \mD_{\mu\nu}V_2^\nu$ reduces to $m_{V_2}^2 h V_1^\mu V_{2\mu}$  for an on-shell $V_2$ boson. This can be seen either by applying the equation of motion for $V_2$ or simply dotting the Feynman vertex with the $V_2$ polarization vector. Therefore, if $V_2=\gamma$, the corresponding operator will not contribute to processes involving an on-shell photon.  However, it was pointed out in Ref.~\cite{Chen:2012jy}, HZ$\gamma$ and H$\gamma\gamma$ couplings contribute non-negligibly to $H\to ZZ^*/Z\gamma^*/\gamma^*\gamma^*\to  4\ell$ channels, which could be leveraged to extract these couplings \cite{Chen:2014gka}. Such analyses have been performed at the LHC \cite{Sirunyan:2017tqd} and the resulting constraints on anomalous HVV couplings are still rather weak. For example, at the $1\sigma$ level various anomalous HVV couplings could still contribute up to ${\cal O}(50\%)$ of the observed HWW and HZZ signal strengths.

\begin{figure}[t]
\centering
  \subfloat[$\Delta\chi^2$ fit on $\xi$ using combined $\kappa_W$ and $\kappa_Z$ from the LHC.]{\includegraphics[width=0.45\textwidth]{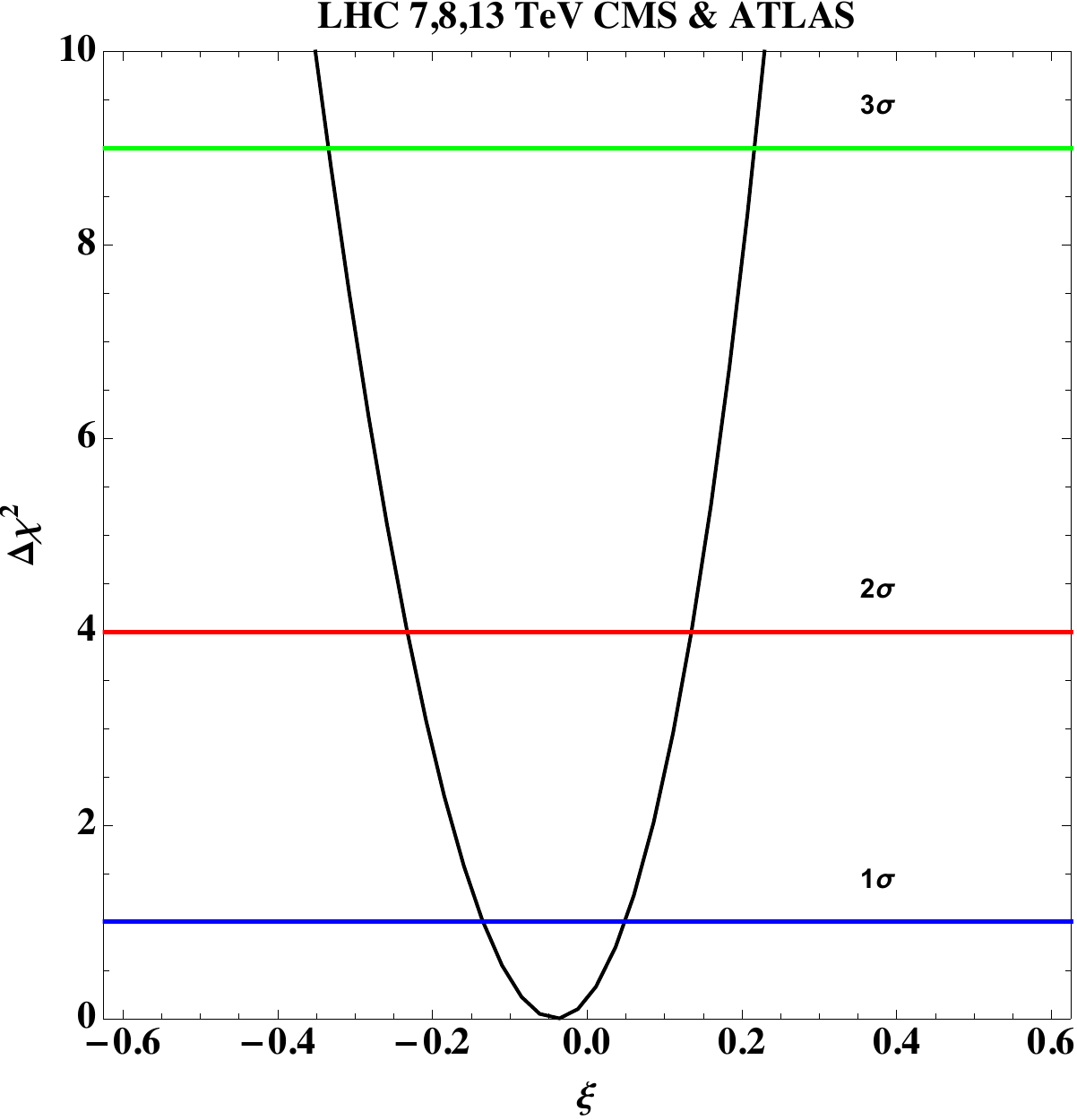}\label{fig:xibound}}
  \quad \ \ 
  \subfloat[$1\sigma$ and $2\sigma$  contours of $\kappa_W$ vs $\kappa_Z$ on $\xi$.]{\includegraphics[width=0.45\textwidth]{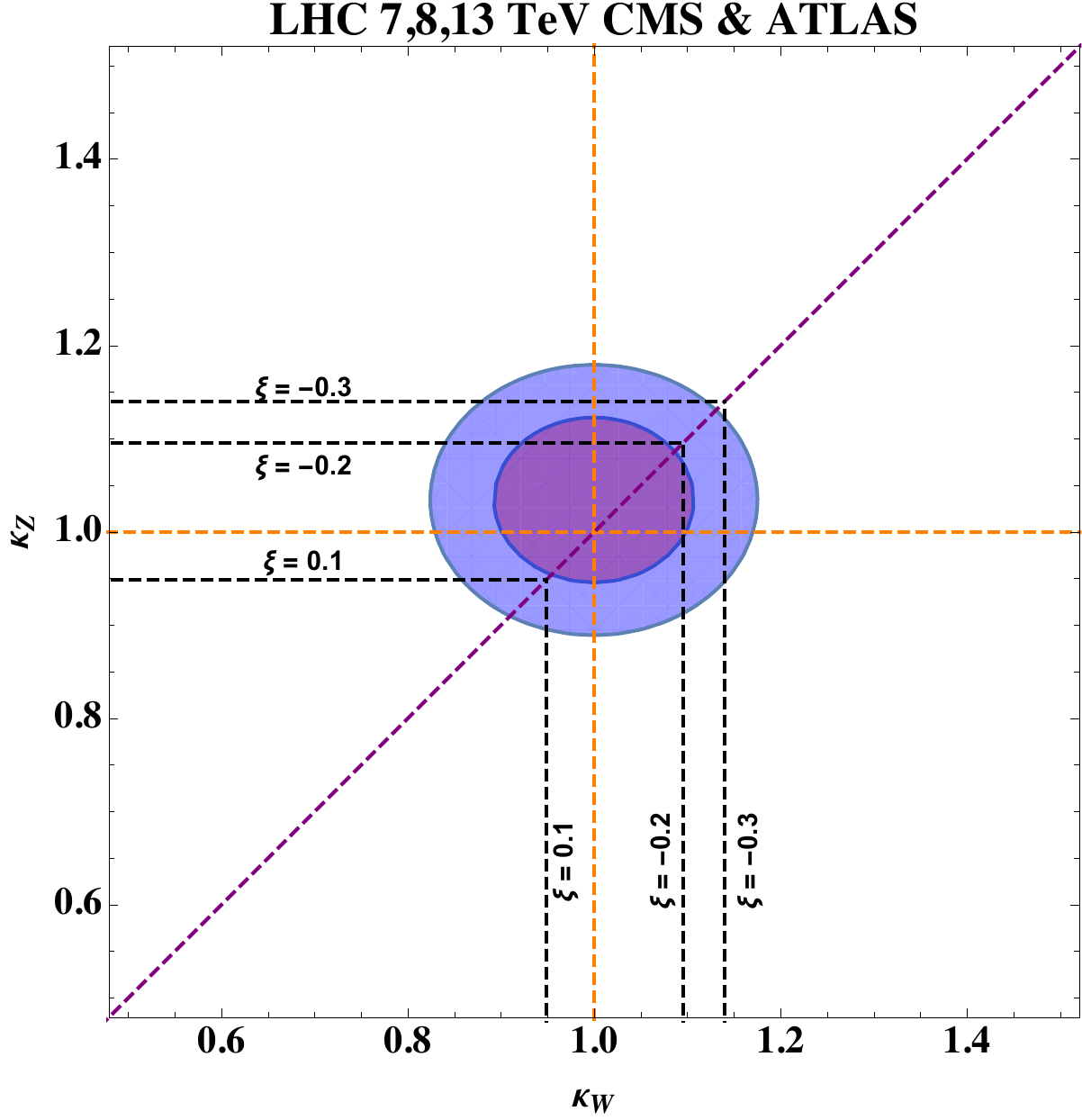}\label{fig:cwcz}}
  \caption{Present model-independent bound on $\xi$ from HVV signal strength measurements at the LHC. From the low-energy perspective, $\xi$ could be either positive (for a compact coset $G/H$) or negative (for a non-compact coset $G/H$).}
  \label{fig:fbound}
\end{figure}

Given that all the universal relations are controlled by one input parameter: the Goldstone decay constant $f$, it is important to have a precise measurement of $f$. Conventionally $f$ is extracted from the signal strengths, $\kappa_W$ and $\kappa_Z$, in $h\to WW$ and $h\to ZZ$ channels using Eqs.~(\ref{eq:L2exp}) and (\ref{eq:notest1}).\footnote{There are constraints on $f$ from considering the fermion sector of a particular composite Higgs model. But these are model-dependent. On the other hand, constraints derived from HVV couplings are independent of $G/H$ and the embedding of the fermion sector at the tree-level \cite{Liu:2018vel}.} For a precise determination of $f$ this is unsatisfactory because ${\cal O}(p^4)$ operators listed in Table~\ref{tab:frhvv} enter into $\kappa_W$ and $\kappa_Z$ as well. A careful analysis including these effects is currently lacking and beyond the scope of the present work. Instead, in Figure~\ref{fig:fbound} we will be content with an experimental constraint on $f$ without including effects of ${\cal O}(p^4)$ operators, by using currently available data at the LHC \cite{Khachatryan:2016vau,ATLAS:2018doi,CMS:2018lkl}. One can see that the bound on $\xi$ is still rather weak, 
\be
\xi=-0.041^{+0.090}_{-0.094}\ [-0.23,0.13]\ ,
\ee
where we show both the allowed 68\% CL (central values with uncertainties) and 95\% CL (in square brackets) intervals. There is a slight preference for a negative $\xi$. It has been pointed out that a positive $\xi$ signals a compact coset $G/H$, while a negative $\xi$ requires a non-compact coset \cite{Low:2014nga,Low:2014oga}. 

In the following we will focus on UR1 in \Eq{eq:ident1}, which involves only the TGC and HVV couplings, and study its potential observability at future lepton colliders. Here we  adopt the viewpoint that the universal relations can be thought of as different ways to measure the parameter $\xi$.  If the $\xi$ extracted from various universal relations all converge on a common value within the experimental precision, that would serve as a smoking gun signal of the ``nonlinear shift symmetry" enforcing the universal relations. Conversely, if the $\xi$ obtained from different universal relations are inconsistent with each other, then the composite Higgs model is falsified.

\begin{table}[t]
\begin{center}
\begin{tabular}{|l|c |c| c| c| c|c|c|}
\hline
  $\mI^{h}_i$ &   $\frac{m_W^2}{m_\rho^2}C^h_i$(ILC)\\
 \hline
(1) $  {h} Z_{\mu} \ewd^{\mu \nu} Z_{\nu}/v$ &$5.83\times 10^{-4}$ \\
 \hline
(2)  $ {h}  Z_{\mu\nu} Z^{\mu\nu}/v$ & $3.93\times 10^{-4}$\\
 \hline
(3)  $ {h}  Z_{\mu} \ewd^{\mu \nu} A_{\nu}/v$ & $$ \\
\hline
(4) $ {h} Z_{\mu\nu} A^{\mu\nu}/v$ & $3.88\times 10^{-4}$\\
\hline
\hline
  $\mI^{3V}_i$ &   $\frac{m_W^2}{m_\rho^2}C_i^{3V}$(ILC)  \\
 \hline
$(\delta g_1^Z)  i\, g\, c_{w} W^{+\mu \nu}  W^-_\mu  Z_\nu  + h.c.$ & $6.1\times 10^{-4} $  \\
\hline
$ (\delta \kappa_\gamma)  i\, e\, W^{+}_{\mu} W^{-}_{ \nu} A^{\mu \nu}  $& $6.4\times 10^{-4} $  \\
\hline
\end{tabular}
\end{center}
\caption{Prospective $1\sigma$ uncertainty at the future lepton colliders, taken from Ref.~\cite{Durieux:2017rsg}.}
\label{tab:1sigmabound}
\end{table}

\begin{figure}[t]
\centering
  \subfloat[Measured $\xi$ as a function of $\delta\kappa_\gamma$, using $N_{\rm UR1}$ as an input.\label{fig:kgammaA}]{\includegraphics[width=0.45\textwidth]{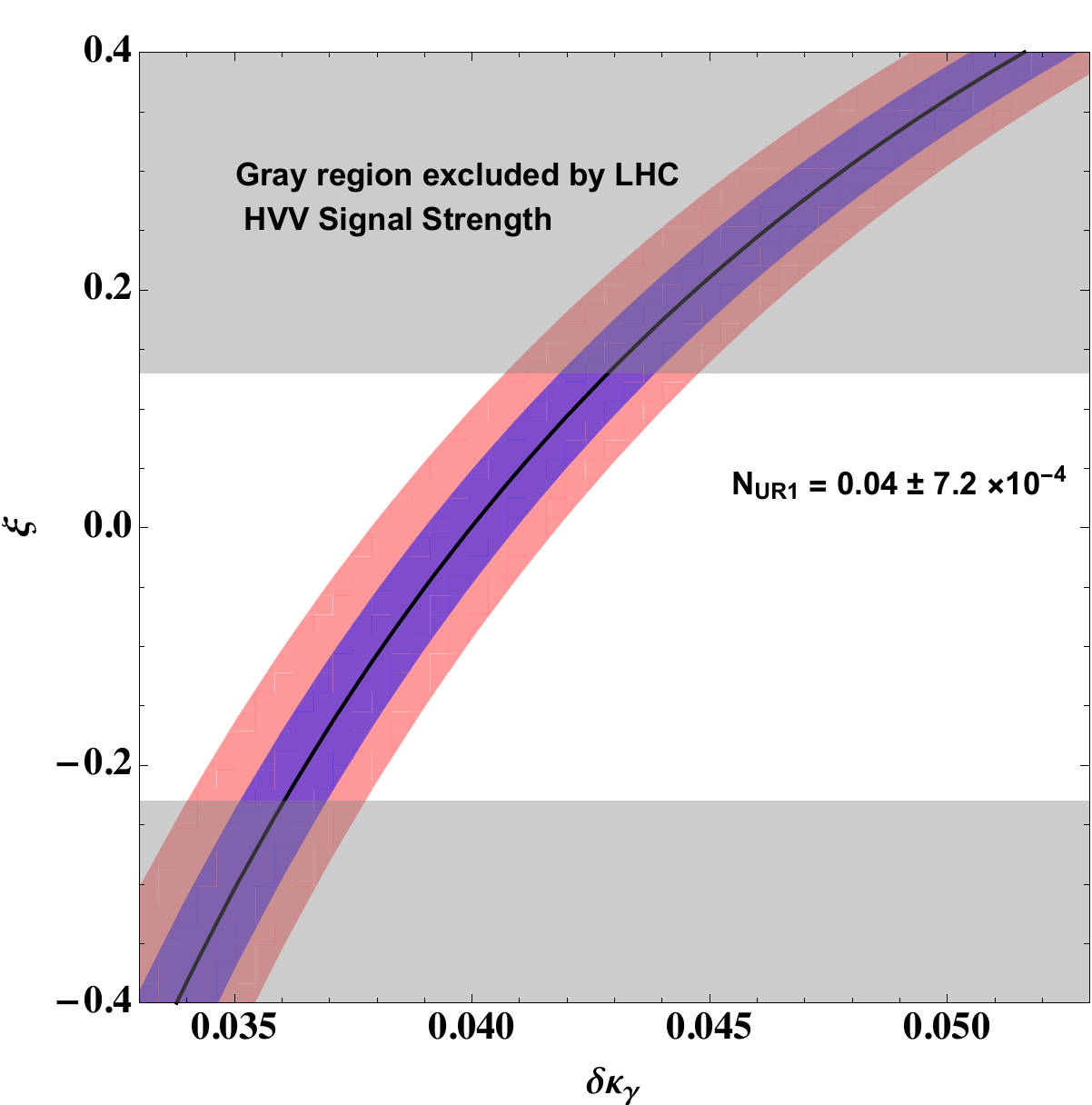}}
  \quad \ \ 
  \subfloat[Measured $\xi$ as a function of $N_{\rm UR1}$, using  $\delta\kappa_\gamma$ as an input.\label{fig:kgammaB}]{\includegraphics[width=0.47\textwidth]{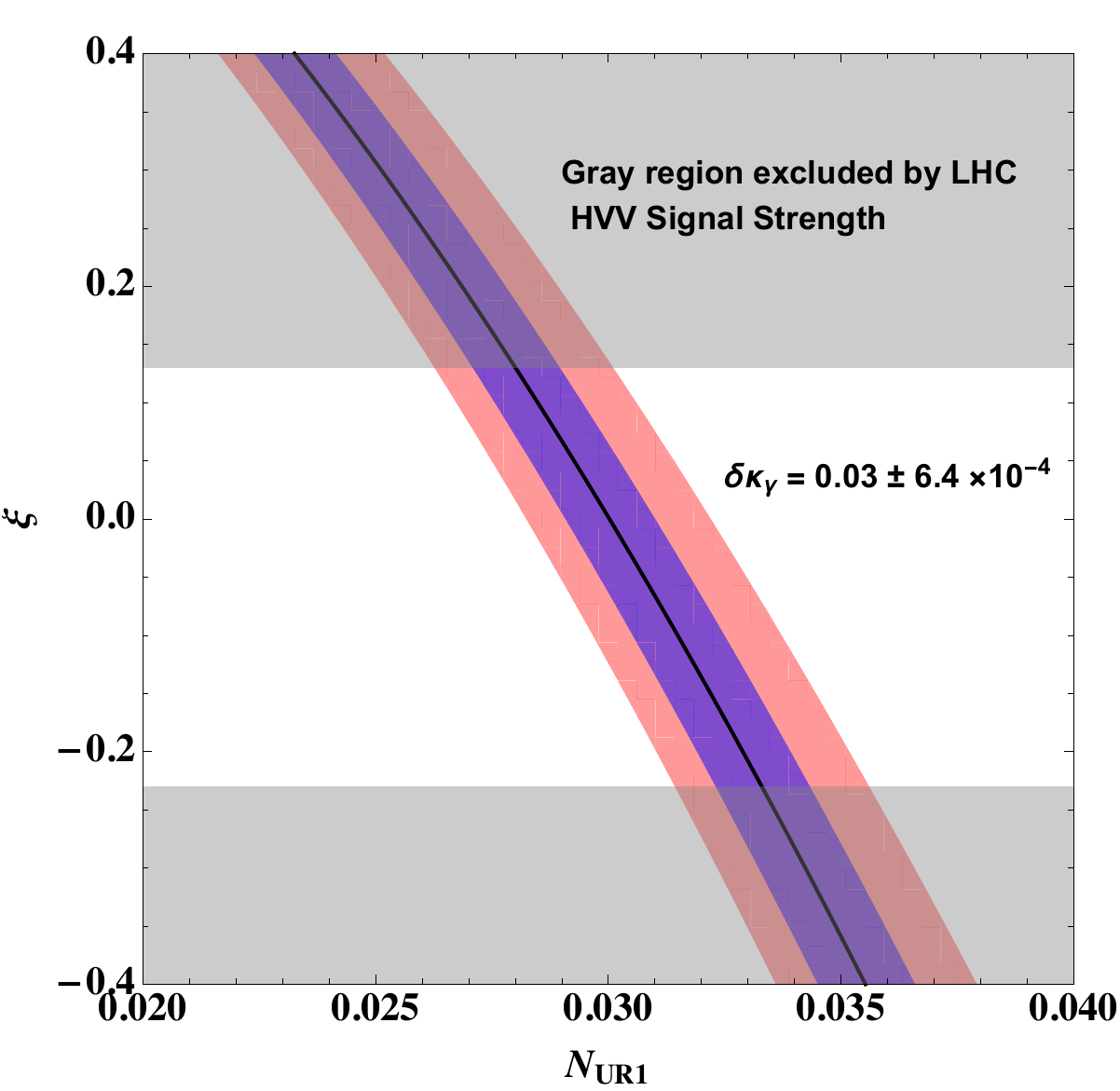}}
  \caption{Using UR1 to measure $\xi$, where the blue and red bands correspond to the $1\sigma$ and $2\sigma$ region on the measurement. \label{fig:kgamma}}

\end{figure}

Our analysis is based on the projection in \Ref{Durieux:2017rsg}, as well as  similar studies in Refs.~\cite{Craig:2015wwr,Gu:2017ckc,Chiu:2017yrx,Ge:2016zro}. The normalization of Wilson coefficients in Table~\ref{tab:oneh}  is different from that defined in Eq.~(A.2) of \Ref{Durieux:2017rsg}.  They are related as follows:
\beq
\begin{split}
\frac{m_W^2}{m_\rho^2}C_1^h &= g^2 c_{Z\Box} \sim 0.42  c_{Z\Box}\ , \qquad  \frac{m_W^2}{m_\rho^2} C_2^h = \frac{g^2+g^{\prime 2}}{4} c_{ZZ} \sim 0.14 c_{ZZ}\ , \\
 \frac{m_W^2}{m_\rho^2} C_3^h &= g g^\prime c_{\gamma\Box} \sim 0.23 c_{\gamma\Box}\ , \qquad  \frac{m_W^2}{m_\rho^2}C_4^h = \frac{e\sqrt{g^2 + g^{\prime 2}}}{2}c_{Z\gamma} \sim 0.12 c_{Z\gamma} \ .
\end{split}
\eeq
One can then obtain the projected precision of various Wilson coefficients in future lepton colliders, which we give in Table \ref{tab:1sigmabound}. The expected $1\sigma$ uncertainty on the coefficients entering UR1 is  
\beq
 N_{\rm UR1}\equiv \frac{m_W^2}{m_\rho^2}\left(2c^2_{w}  C^{h}_2 -c_{2w} C^{h}_4/t_{w}\right) : 7.20 \times 10^{-4}\ ,\quad
\delta{\kappa}_\gamma : 6.4\times 10^{-4} \ ,
\eeq
Then UR1 in \Eq{eq:ident1} can be written as\footnote{Since we are taking the ratio of Wilson coefficients, the particular normalization chosen in Eq.~(\ref{eq:Cis}) becomes irrelevant.}
\be
{\rm UR1}: \frac{N_{\rm UR1}}{\delta\kappa_\gamma} = \sqrt{1-\xi} \ .
\ee
Both $N_{\rm UR1}$ and $\delta\kappa_\delta$ can be extracted experimentally. In Fig.~\ref{fig:kgammaA} we show the measured $\xi$ and its uncertainty as a function of $\delta\kappa_\gamma$, using a benchmark $N_{\rm UR1}$ as the input, while in Fig.~\ref{fig:kgammaB} we reverse the roles of $\delta\kappa_\gamma$ and $N_{\rm UR1}$. The gray area is excluded by current HVV signal strength measurements at the LHC, and the allowed region is still quite large today. However, in a future lepton collider the precision on HVV signal strength is expected to be at the sub-percent level \cite{Durieux:2017rsg},\footnote{At that level of precision, one should perform a global fit of $\xi$ by including ${\cal O}(p^4)$ operators in the HVV coupling measurements.} which would allow for a precise determination of $\xi$ that is independent of UR1. If the $\xi$ extracted from two separate channels turn out to agree with each other within experimental uncertainty, it would constitute a striking confirmation of the underlying shift symmetry acting on the 125 GeV Higgs boson. Otherwise, a generic composite Higgs boson would be strongly disfavored.

\section{Conclusion and Outlook}

In this work we presented a number of universal relations in models where the Higgs arises as a pseudo-Nambu-Goldstone boson. These relations are dictated by the underlying nonlinearly realized symmetry acting on the 125 GeV Higgs boson, which is embodied in the nonlinear shift symmetry in Eq.~(\ref{eqshift}). From the infrared perspective, the shift symmetry simply enforces  the correct single soft limit in the S-matrix elements of the Nambu-Goldstone boson and turned out to be insensitive to the nature of the broken group $G$ in the UV. Under the well-motivated assumption of the Higgs transforming as the $\mathbf{4}$ of an $SO(4)$ subgroup of the unbroken group $H$, the shift symmetry allows one to construct a universal effective Lagrangian for the composite Higgs boson, without reference to a particular symmetry breaking pattern $G/H$. In particular, interactions of the 125 GeV Higgs with the electroweak gauge bosons: the HVV, HHVV, HVVV and TGC couplings remain universal even after integrating out other heavy composite resonances \cite{Liu:2018vel}. Universal relations are ratios among the HVV, HHVV, HVVV and TGC couplings that depend on only one input parameter: the Goldstone decay constant $f$. Experimental verification of the universal relation would constitute a coset-independent smoking gun signal of the pseudo-Nambu-Goldstone nature of the Higgs boson.

We presented the complete list of predictions from a composite Higgs boson in HVV, HHVV, HVVV and TGC couplings to all orders in $1/f$ and up to ${\cal O}(p^4)$ in Tables \ref{tab:oneh} - \ref{tab:tgc}, as well as a number of universal relations, which are expressed in terms of coefficients   that can be measured experimentally. These coefficients involve different tensor structures in HVV, HHVV, HVVV and TGC couplings. To facilitate future phenomenological analyses, we provided the Feynman rules in Table \ref{tab:frhvv} - \ref{tab:frhhvv}.

Because the universal relations all involve one single input parameter $f$, they can be viewed as different ways to extract $f$ experimentally. A composite Higgs boson would then manifest itself in the consistent measurement of $f$ from different universal relations. Conversely, if different measurements arrive at incompatible values of $f$, it would invalidate the nonlinear symmetry acting on the composite Higgs boson. 

As a preliminary study, we provided an updated bound on $f$ using the signal strength measurement on HVV couplings, without including $O(p^4)$ effects in the effective Lagrangian, as is conventionally done in the literature. In the future it would be desirable to include these effects for a precise determination of $f$.  Then we proceeded to study UR1 in Eq.~(\ref{eq:ident1}), which involves only the HVV and TGC couplings at future lepton colliders. We presented the expected precision on the extraction of $\xi$ using UR1 in Fig.~\ref{fig:kgamma}.

\begin{figure}[tbp]
\centering
  \subfloat[Double Higgs production through vector boson fusion at a hadron collider.]{\includegraphics[height=4.4cm]{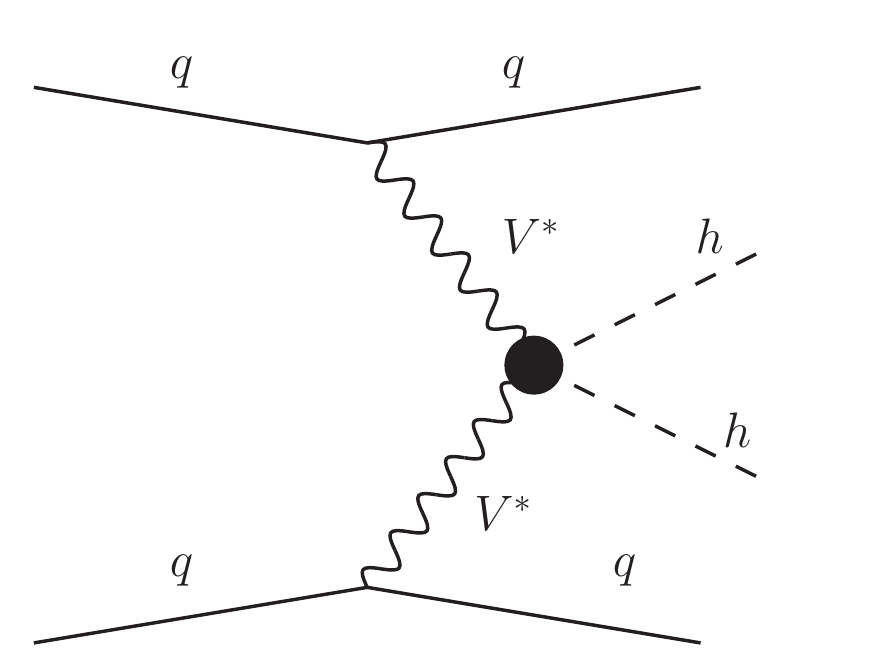}}
  \quad \ \ 
    \subfloat[Double Higgs production through vector boson fusion at a lepton collider.]{\includegraphics[height=4.4cm]{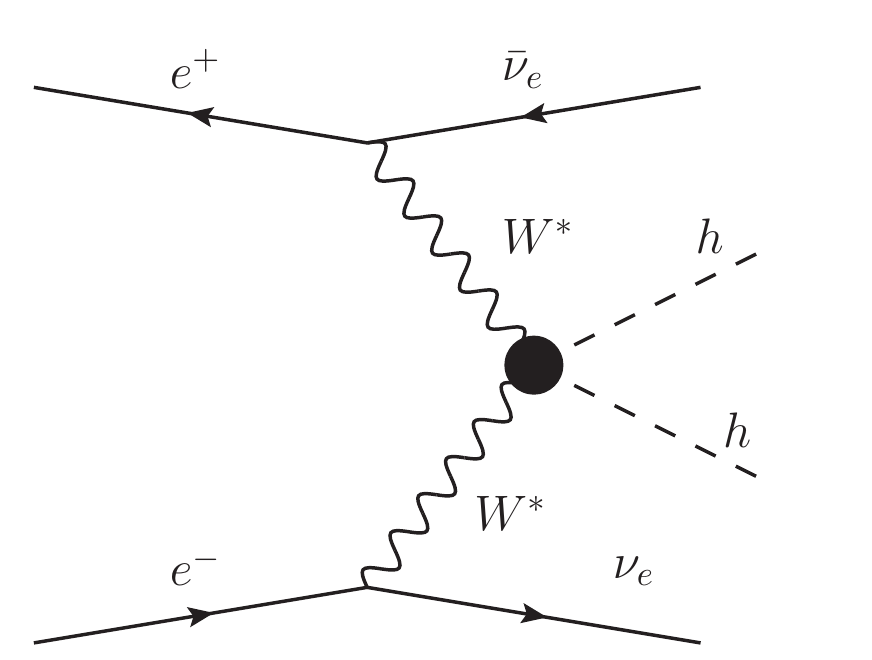}}
  \quad \ \ 
  \subfloat[Double Higgs production in association with a vector boson.]{\includegraphics[height=3.3cm]{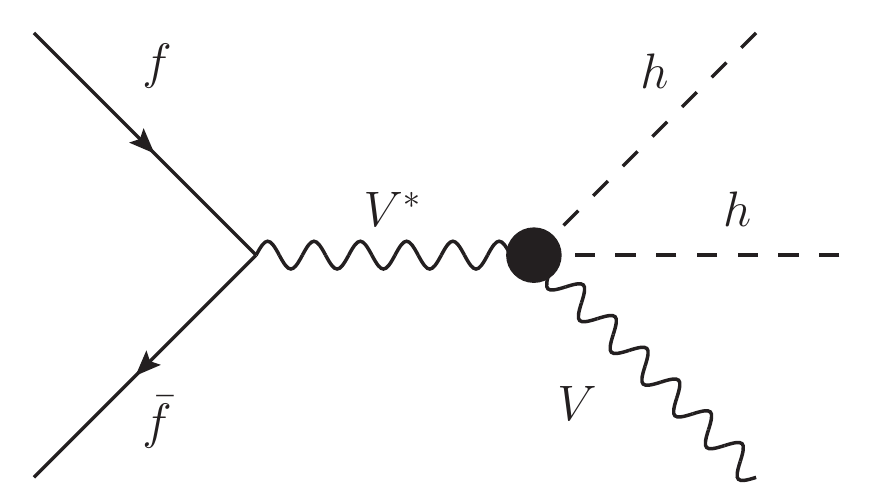}}
  \quad \ \ 
  \subfloat[Off-shell Single Higgs decay.]{\includegraphics[height=3.3cm]{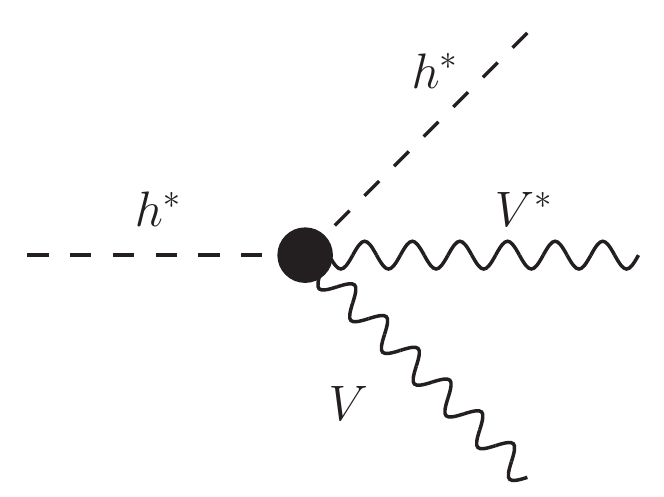}}
  \caption{Production and decay topology of venues to test the HHVV couplings. A black dot represents contributions from various Feynman diagrams. To measure the HVVV coupling, replace one of the external Higgs particle by an electroweak gauge boson.\label{fig:testhh}}
\end{figure}

Last but not least, we comment on the prospect of testing universal relations involving HHVV and HVVV couplings. In particular, the HHVV coupling is predicted in the SM but among the least studied experimentally. For this reason alone, studying HHVV coupling should be among the top priority in future experimental programs on the Higgs boson. We see the following venues to test the HHVV coupling, which are shown in Fig. \ref{fig:testhh}:
\begin{enumerate}[label=(\alph*)]

\item Double Higgs production through vector boson fusion (VBF) in a hadron collider: $qq \to 2h+2j$.

\item Double Higgs production through vector boson fusion (VBF) in a lepton collider: $e^+e^-\to 2h+\nu_e\bar{\nu}_e$.

\item Double Higgs production  in association with a vector boson: $f\bar{f}\to V^*\to 2h+V$. In a hadron collider the initial states are quarks and in an $e^+e^-$ collider they are electrons.

\item Off-shell single Higgs decay: $h^*\to h^*V^*V$.

\end{enumerate}
The VBF channel  has been studied previously \cite{Dolan:2013rja,Dolan:2015zja,Bishara:2016kjn,Arganda:2018ftn,Kilian:2018bhs} and the prospect at a future lepton collider has also been considered \cite{Contino:2013gna,Kanemura:2016tan,CLIC:2016zwp,Barklow:2017awn,DiVita:2017vrr}. On the other hand, we are not aware of any studies in the associated production channel or the off-shell single Higgs decay channel. More importantly, as have been emphasized repeatedly, the ultimate goal is to not only measure the signal strength in these channels, but also test the specific tensor structures predicted in the SM, in much the same way we verify the HVV coupling at the Run 1 of the LHC. This aspect of testing the HHVV structure has received very little attention in current literature. It is also clear that the same production and decay topology can be used to measure HVVV couplings, by replacing one of the external Higgs bosons by an electroweak gauge boson. Obviously, these are very challenging experimental tasks and it is desirable to introduce advanced  tools to facilitate the  analysis. Much remains to be done, and we hope to return to some of these topics in the future.

\begin{acknowledgments}
 
 D.L. and Z.Y. acknowledge useful discussions with Jiayin Gu, Zhen Liu and Jing Shu.  D.L. and I.L. acknowledge the hospitality of CERN Theory Department where part of this work was performed. D.L. would also like to thank the Particle Phenomenology Group at EPFL for the generous support. This work is supported in part by the U.S. Department of Energy under contracts No. DE-AC02-06CH11357 and No. DE-SC0010143.
  \end{acknowledgments}

\appendix

\section{Generators of $SO(4)$}
\label{appen:so5}

The expression for the fundamental representation of $SO(4)$ generators is reported as follows:
\beq
\begin{split}
T^{r L/R}_{ab} &= -\frac{i}{2} \left[ \frac12 \epsilon^{rst}(\delta^{sa}\delta^{tb} -\delta^{sb}\delta^{ta} )\pm (\delta^{ra}\delta^{4b} - \delta^{rb} \delta^{4a}  )\right]\,.\\
\end{split}
\eeq
where $a= 1,2,3,4$ and $r,s,t = 1,2,3$. The generators are normalized as $\text{Tr}[T^iT^j] = \delta ^{ij}$. $T^{rL/R}$ are the generators satisfying  $SU(2)$ Lie-algebra:
\beq
\begin{split}
[T^{rL}, T^{sL}] &= i \epsilon^{rst} T^{tL},\qquad [T^{rR}, T^{sR}] = i \epsilon^{rst} T^{tR},\qquad [T^{rL}, T^{sR}] = 0,
\end{split}
\eeq

\section{Matching to Dimension-6 Operators in SMEFT}
\label{app:d6}

\begin{table}[t]
\begin{center}
\footnotesize
\begin{tabular}{|c|c|c|c|c|c|c|c|c|c|}
\hline
 $ f^2\mO_3$ &    $ f^2\mO_4^+$ &   $ f^2\mO_4^-$ \\
\hline
  \textcolor{blue}{ $-4 (\mO_W - \mO_B)$ }&  \textcolor{blue}{$2(\mO_{HW} + \mO_{HB})$} &\textcolor{blue}{ $ 2(\mO_{HW} - \mO_{HB})$} \\
  \hline \hline
  $f^2\mO_5^+$ &  $f^2\mO_5^-$  &   \\
\hline
\textcolor{blue}{$ 4\left[\mO_{W} + \mO_{B} - (\mO_{HW} + \mO_{HB}) \right]$ } & \textcolor{blue}{ $ -4 \left[\mO_{W} - \mO_{B} - (\mO_{HW} - \mO_{HB})\right]$} &   \\
\hline
\end{tabular}
\end{center} 
\caption{ Matching of the $\mO(p^4)$ opeartors in \Eq{eq:op4} to  the dimension-6 operators defined in \Eq{eq:d6ops}.}
\label{tab:LOMatching}
\end{table}

In this appendix, we present the matching of the universal nonlinear Lagrangian to the SMEFT at the dimension-6 order. We will use the SILH basis defined  in \Ref{Giudice:2007fh} and the operators relevant for this calculation are as follows:
\beq
\small
\begin{split}
&\mO_H = \frac{1}{2}\partial_\mu (H^\dagger H)  \partial^\mu (H^\dagger H),\quad {\cal O}_T = \frac{1}{2}(H^\dagger \olr{D}^\mu H )(H^\dagger \olr{D}_\mu H) \\
& \mO_6 =\lambda (H^\dagger H)^3, \quad \mO_{y} = y_f H^\dagger H \bar{f}_L H f_R  \\
&{\cal O}_W =\frac{ig}{2 }\left( H^\dagger  \sigma^a \overleftrightarrow{D}^\mu H \right )D^\nu  W_{\mu \nu}^a , \qquad 
{\cal O}_B =\frac{ig'}{2 }\left( H^\dagger  \overleftrightarrow{D}^\mu H \right )\partial^\nu  B_{\mu \nu} \\
&{\cal O}_{HW} = i g(D^\mu H)^\dagger\sigma^a(D^\nu H)W^a_{\mu\nu}, \qquad
{\cal O}_{HB}= i g'(D^\mu H)^\dagger(D^\nu H)B_{\mu\nu} \ .
\end{split}
\label{eq:d6ops}
\eeq
The  effective Lagrangian for the dimension-six operators is parameterized as:
\beq
\label{eq:d6lag}
\mL^{\text{(D6)}} = \sum_{i = H, T,y,6} \frac{c_i}{f^2} \mO^{(6)}_i + \sum_{i=W,B,HW,HB} \frac{c_i}{m_{\rho}^2} \mO^{(6)}_i\ ,
\eeq
where  we have used slightly different normalization from SILH  for the $\mO_{HW,HB}$.\footnote{The SILH assumes minimal coupling for gauge fields and inserts a one-loop factor for $\mO_{HW,HB}$.}  The leading order $\mO(p^2)$  Lagrangian in \Eq{eqnlsmlagep} gives a contribution to $\mO_H,\mO_y$ and  $\mO_6$ whose Wilson coefficients are completely fixed by the universal nonlinearity 
\beq
\label{eq:op2match}
c_H = 1, \qquad c_y = -\frac{1}{3}, \qquad c_6 = - \frac43\ .
\eeq
One may wonder about the appearance of non-derivatively coupled operators in $\mO_y$ and $\mO_6$ from matching to the two-derivative Lagrangian. This is because, in the SILH operator basis the operator ${\cal O}_r=H^\dagger H D_\mu H^\dagger D^\mu H$, which is present in \Eq{eqnlsmlagep}, is eliminated by a field redefinition, thereby giving rise to  the following operator identity,
\beq
H^\dagger H D_\mu H^\dagger D^\mu H= \frac12\left( \mO_{y_u} + \mO_{y_d}  +  \mO_{y_e} \right)-\mO_H + 2 \mO_6  \ .
\eeq
Eq.~(\ref{eq:op2match}) follows from this identity.

For the  $\mO(p^4)$ operators in \Eq{eq:op4}, we  list their matching to the dimension-6 operators in Table~\ref{tab:LOMatching}. Note that for the four-derivative operators $\mO_1$ and $\mO_2$, the leading contribution to the matching appears at the dimension-8, which is why they do not appear in  Table~\ref{tab:LOMatching}. This is also the case for one linear combination of the remaining five operators $\mO_{3}$, $\mO_4^\pm$ and  $\mO_{5}^\pm$. As a result,  these five operators give rise to  only four operators in SMEFT: ${\cal O}_{W,B,HW,HB}$.

In the unitary gauge, the four operators $\mO_{W,B,HW,HB}$ become
\bea
\label{eq:OWB}
\mO_{W}&=&  2 m_W^2\left( \frac{h}{v}+ \frac{h^2}{2 v^2} \right) \Big[W_\mu^- \ewd^{\mu \nu } W_\nu^+ + W_\mu^+ \ewd^{\mu \nu } W_\nu^-  +  Z_\mu \ewd^{\mu \nu}  Z_{\nu}  +  t_w Z_{\mu} \ewd^{\mu \nu} A_{\nu} \Big] \nonumber\\
 &&-\frac{t_w}{2} m_W^2  W_{\mu\nu}^{(3)}B^{\mu\nu} - \frac{i g m_W^2  }{  c_w}\left[ ( W^+_{\mu \nu}  W^{- \mu} -  W^-_{\mu \nu} W^{+\mu})  Z^\nu + Z^{\mu \nu} W^+_\mu W^-_{\nu} \right] \nonumber\\
&&- 2i m_W^2\left( \frac{h}{v}+ \frac{h^2}{2 v^2} \right) \left[ \left(\frac{g}{c_w} (1 + 2 c_w^2) Z^\nu + 2 e A^{\nu}\right) \left(W_{\mu\nu}^+ W^{\mu-} - W_{\mu\nu}^- W^{\mu+}  \right)\right. \nonumber\\
&&\left.\qquad+\left(\frac{g}{c_w} (1 + 2 c_w^2)Z^{\mu\nu} + 2 e A^{\mu\nu}\right) W_{\mu}^+ W_{\nu}^- \right] \nonumber\\
&&- 2i g^\prime m_W^2\left(1 + \frac h v\right)\frac{\partial^\nu h}{v} B^\mu\Big[ W_{\mu}^+ W_{\nu}^- -W_{\mu}^- W_{\nu}^+\Big]-\frac{g^2}{2}m_W^2   (W_{\mu}^+ W_{\nu}^- -W_{\mu}^- W_{\nu}^+ )^2 \nonumber\\
&&+ 2g^2 m_W^2 \left( \frac{h}{v}+ \frac{h^2}{2 v^2} \right) \left(W_\mu^a  - t_w \delta^{a3} B_\mu\right) \left(W^{\mu a} (W_\nu^b)^2 - W^{\mu b} W_\nu^a W^{\nu b}\right) \nonumber\\
&& + \frac{2m_W^2 g^2}{c_w} Z_\nu W^{\nu 3} W_\mu^+ W^{\mu -}  - \frac{m_W^2 g^2}{c_w} Z^\nu W^{\mu 3} \left(W_\mu^+ W_\nu^{ -}  + W_\mu^- W_\nu^{ +} \right) \ , \\
\mO_{B}  &=&   -\frac{t_w}{2}m_W^2 W^{(3)}_{\mu\nu}B^{\mu\nu} - \frac{2t_w}{c_w} m_W^2 \left(\frac{h}{v} + \frac{h^2}{2 v^2}\right) \left(-s_w Z_\mu \mD^{\mu\nu} Z_\nu + c_w Z_\mu \mD^{\mu\nu} A_\nu \right)\  ,
\eea
\bea
\label{eq:OHWHB}
\mO_{HW} &=&  2 m_W^2\left( \frac{h}{v}+ \frac{h^2}{2 v^2} \right) \Big[W_\mu^- \ewd^{\mu \nu } W_\nu^+ + W_\mu^+ \ewd^{\mu \nu } W_\nu^-  +  Z_\mu \ewd^{\mu \nu}  Z_{\nu}  +  t_w Z_{\mu} \ewd^{\mu \nu} A_{\nu}  \nonumber\\
&& \qquad- W^{+ \mu \nu} W^-_{\mu \nu}- \frac12 Z_{\mu \nu}^2- \frac{t_w}{2}  Z^{\mu \nu} A_{\mu \nu}\Big] \nonumber \\
&& - \frac{2 i g m_W^2  }{  c_w}  \left(\frac12 + \frac h v+\frac{h^2}{2v^2}\right)\left[ ( W^+_{\mu \nu}  W^{- \mu} -  W^-_{\mu \nu} W^{+\mu})  Z^\nu \right.\nonumber\\
&&\qquad \left.+ (c_w^2 Z^{\mu \nu} + s_w c_w  A^{\mu \nu})W^+_\mu W^-_{\nu} \right] \nonumber\\
&&+ 4 g^2 m_W^2  \left(\frac12 + \frac h v+\frac{h^2}{2v^2}\right) \left(W_\mu^- W_\nu^+ W^+_{[\mu}W^-_{\nu]} + \frac{1}{c_w}W^-_\mu Z_\nu W^+_{[\mu}W^3_{\nu]} \right.\nonumber\\
&&\qquad \left.+ \frac{1}{c_w}W^+_\mu Z_\nu W^-_{[\mu}W^3_{\nu]}\right) \nonumber \\
&&-2i g^\prime m_W^2 \left(1 + \frac h v\right) \frac{\partial_\nu h}{v} B^\mu\left( W^+_{\mu} W^-_{\nu} -  W^-_{\mu} W^+_{\nu}  \right) \ ,\\
\mO_{HB} &=& 2 m_W^2\left( \frac{h}{v}+ \frac{h^2}{2 v^2} \right) \left[  t_w^2 Z_\mu \ewd^{\mu \nu}  Z_{\nu} - t_w   Z_{\mu} \ewd^{\mu \nu} A_{\nu}-\frac{t^2_w}{2}   (Z_{\mu \nu})^2  + \frac{ t_w}{2}Z^{\mu \nu} A_{\mu \nu} \right] \nonumber \\
&&- 2 i g t_w m_W^2 \left(\frac12 + \frac h v+\frac{h^2}{2v^2}\right)  (-s _w Z^{\mu \nu} +  c_w  A^{\mu \nu})W^+_\mu W^-_{\nu} \ .
\eea
Note that $\mO_{W,B}$ will contribute to the propagators of  $W,B$ gauge bosons and one should canonically normalize the gauge boson kinetic terms in order to obtain the physical couplings. To be more specific, in the unitary gauge, $\mO_W$ can be written as:
\beq
\mO_W = \frac{g^2}{4} (h + v)^2 \left(W^{a\mu} - t_w \delta^{a 3} B^\mu\right) D^\nu W_{\mu\nu}^a \ .
\eeq
Upon integration-by-parts, one can obtain a term proportional to $W_{\mu\nu}^a W^{a\mu\nu}$, which should be absorbed into the definition of the gauge couplings by canonically normalizing gauge kinetic term. Similar reasoning applies to $\mO_B$ as well.
In \Eq{eq:OWB} and \Eq{eq:OHWHB}, we have performed such redefinitions of the gauge  couplings.

\bibliography{references_Higgs}
\bibliographystyle{JHEP}

\end{document}